\documentclass[a4paper,11pt]{article}

\usepackage{color}

\usepackage{fullpage}
\usepackage[T1]{fontenc}

\newcommand{\be}{\begin{eqnarray}}
\newcommand{\ee}{\end{eqnarray}}

\usepackage{amssymb,amsmath,amsfonts,amsthm}
\usepackage{bbold}
\usepackage{hyperref}
\usepackage{graphicx}
\usepackage{cite}
\numberwithin{equation}{section}
\DeclareMathOperator{\sgn}{sgn}

\begin{document} 
	
\begin{titlepage}
	\thispagestyle{empty}

{\ }
	
	\vspace{35pt}
	
	\begin{center}
	    { \LARGE{\bf On scale-separated supersymmetric AdS$_2$ flux vacua}}
		
		\vspace{50pt}

		{\Large N.~Cribiori$^{1,2}$, F.~Farakos$^{3}$, N.~Liatsos$^{3}$} 

		\vspace{25pt}

		{
			$^1$ {\it  KU Leuven, Institute for Theoretical Physics,  \\ 
        Celestijnenlaan 200D, B-3001 Leuven, Belgium }

		\vspace{15pt}

  $^2$ {\it Max-Planck-Institut f\"ur Physik (Werner-Heisenberg-Institut), \\ 
   Boltzmannstra\ss e 8 ,  85748 Garching, Germany,}

	\vspace{15pt}
 
			$^3${\it Physics Division, National Technical University of Athens \\
        15780 Zografou Campus, Athens, Greece}

		}

		\vspace{40pt}

		{ABSTRACT}
	\end{center}
	
	\vspace{10pt}
	
	We argue that scale-separated AdS$_2$ vacua with at least two preserved supercharges cannot arise from flux compactifications in a regime of computational control. We deduce this by showing that the AdS$_2$ scale is parametrically of the same order as the tension of a fundamental BPS domain wall, which provides an upper bound on the UV cutoff. 
    Since the latter does not need to be associated to any geometric scale, the argument excludes scale separation in a broader sense than what commonly considered. 
    Our claim is exemplified by a bottom-up 2D supergravity analysis as well as top-down models from Type II flux compactifications. 

\bigskip

\end{titlepage}

\tableofcontents

\baselineskip 6 mm

\section{Introduction} 

One of the main open problems in fundamental physics is to explain from first principles why only four macroscopic spacetime dimensions are observed at low energy. This is particularly important if one aims at connecting ten-dimensional superstring theory to the real world. Addressing this problem on a positive background energy density is currently out of reach, since one would also need to understand the mechanism of spontaneous breaking of supersymmetry, which is another major challenge. Hence, a simplification is to consider the less realistic case with negative cosmological constant, namely anti-de Sitter (AdS) vacua. In this context, the problem has been addressed, but not solved, with the following attitude. 
From a top-down perspective, one asks for the existence of a (parametric) hierarchy of scales between the AdS length scale $L$ and the Kaluza-Klein scale \cite{Tsimpis:2012tu,Petrini:2013ika,Gautason:2015tig}. 
From a bottom-up perspective, which is the one adopted here, one asks for a hierarchy between the AdS radius and the ultraviolet cutoff of the effective theory, $\Lambda_{UV}$, namely
\begin{equation}
\label{eq:defscalesep}
\Lambda_{UV} L \gg 1.
\end{equation}
Vacua satisfying this requirement are scale-separated and can explain why only an AdS factor of spacetime is large enough to be visible at low energy. Let us stress that an advantage of this bottom-up approach is that the scale $\Lambda_{UV}$ does not necessarily need to have a geometric interpretation, so that one can also cover more general situations. On the contrary, the Kaluza-Klein scale is typically associated to the geometry of the extra dimensions. 

Along with this logic, a large portion of the parameter space of scale-separated AdS vacua in various dimensions and with different amounts of preserved supercharges has already been ruled out. In particular, in \cite{Cribiori:2022trc,Cribiori:2023ihv} a general argument has been provided showing that AdS vacua with at least eight preserved supercharges at the Lagrangian level are in tension with the weak gravity conjecture \cite{Arkani-Hamed:2006emk}, which is among the most tested properties of quantum gravity. A complementary argument has been presented in \cite{Montero:2022ghl} exploiting the absence of global symmetries, another of the most solid features of quantum gravity. Non-supersymmetric AdS vacua in principle evade these arguments, but they are nevertheless conjectured to be unstable \cite{Ooguri:2016pdq}. Besides excluding scale separation as a consequence of those properties of quantum gravity which are best understood, in \cite{Lust:2019zwm} it has been directly conjectured that scale separation in AdS should not occur unless at infinite distance in moduli space, where the effective description breaks down. Recent progress in connecting this conjecture to more established properties of quantum gravity can be found in \cite{Shiu:2022oti,Li:2023gtt,Basile:2023rvm,Shiu:2023bay,Palti:2024voy,Mohseni:2024njl,Debusschere:2024rmi}.

Since the number of independent components of an irreducible spinor is smaller than eight only in less than five spacetime dimensions, the argument of \cite{Cribiori:2022trc,Cribiori:2023ihv} can be evaded in four or less spacetime dimensions even when preserving some supersymmetry. Indeed, this is the place where proposals for scale-separated vacua in string theory have been put forward. While a recent review can be found in \cite{Coudarchet:2023mfs}, let us nevertheless mention the celebrated DGKT construction in compactifications of massive type IIA supergravity to four dimensions \cite{Derendinger:2004jn,Behrndt:2004mj,DeWolfe:2005uu}, improved in \cite{Junghans:2020acz,Marchesano:2020qvg} beyond the smeared approximation, and the more recent double T-dual version in massless type IIA at both weak and strong coupling (M-theory) \cite{Cribiori:2021djm}. These setups have been further extended and analyzed in \cite{Emelin:2020buq,Conlon:2021cjk,Apers:2022zjx,Andriot:2022yyj,Apers:2022vfp,Andriot:2022brg,Carrasco:2023hta,Tringas:2023vzn,Andriot:2023fss,Emelin:2024vug,VanHemelryck:2024bas}, while three-dimensional type II flux vacua have been obtained in \cite{Farakos:2020phe,Emelin:2022cac,VanHemelryck:2022ynr,Farakos:2023nms,Farakos:2023wps,Arboleya:2024vnp}.

To the best of our knowledge, no scale-separated AdS vacuum in two dimensions (2D) has been constructed so far, but also no explicit argument against it has been provided. Notably, in \cite{Lust:2020npd} a systematic search for scale-separated AdS$_2$ vacua in type IIA compactifications without sources was performed, but no candidate could be identified. In the more recent work \cite{Guarino:2024zgq}, scale separation is achieved between AdS$_2$ and an H$_2$ factor of the compact manifold, but there is still a circle parametrically of the same size as the external space. It is then an important conceptual open problem if 2D gravity in AdS exists on its own, or if it requires the presence of additional macroscopic dimensions to be quantum-mechanically consistent. 
In the present work, we provide a bottom-up explanation of why supersymmetric AdS$_2$ vacua with at least two preserved supercharges cannot satisfy \eqref{eq:defscalesep} and thus are not scale-separated. More precisely, we argue that if a $\mathcal{N}\geq(1,1)$ supersymmetric AdS$_2$ phase of quantum gravity supported by fluxes exists, it must necessarily be higher dimensional. Our argument relies on the assumption that there exists a supersymmetric domain wall separating the 2D background spacetime into two AdS$_2$ domains.

As is well known, gravity in 2D is qualitatively different than in higher dimensions. This implies that we should be careful when trying to extrapolate properties of quantum gravity to 2D, such as the weak gravity conjecture or the absence of global symmetries, in order to constrain scale separation. Crucially, our argument is independent of any of these properties and in fact it cannot be generalized straightforwardly to higher dimensions. This is perhaps to be expected for a genuine quantum gravity statement. Besides, our argument is purely bottom-up and therefore also independent of the details of any microscopic completion. In particular, it does not assume the existence of a geometric (Kaluza-Klein) scale, and it can thus be applied to setups in which the latter is not present, but still the notion of scale separation can be made sense of.

\section{The main argument} 
\label{sec:arg}

In this section we present our main argument excluding scale-separated AdS$_2$ flux vacua. A number of physical assumptions are entering it, and for the sake of clarity we highlight them, but we will see that most of them are common properties of flux compactifications (or of typical gravitational systems) and as such are not stringent.

Let us start by recalling that in 2D the cosmological constant $\Lambda$ is related to the AdS$_2$ length-scale $L$ by
\begin{equation}
\label{eq:RLAdS2}
\Lambda = - \frac{2}{L^2} \,. 
\end{equation}
Crucially, the 2D Planck scale $M_P$ does not appear in the above expression. This property is special to 2D gravity and it is the reason why our analysis cannot be generalized to other dimensions, as we will explain in a while. 
Our main working hypothesis is that we consider AdS$_2$ flux vacua. 
These are such that their curvature is completely supported by (electric) gauge field strengths along the AdS$_2$ directions, 
\begin{equation}
\label{AdSL}
- \frac{2}{L^2} =  \frac{1}{2} |F_2|^2 \,,
\end{equation}
where $|F_2|^2 \equiv \frac{1}{2} F_{\mu \nu} F^{\mu \nu}$.\footnote{Recall that the gauge coupling of a $p$-form field strength in $D$ dimensions has mass dimension $[g^2]=2p-D$. We absorb it inside the gauge fields, such that the $p$-form field strength and the charges have mass dimensions $[F_p]=D/2=[Q]$. The tension of a domain wall has mass dimension $[T]=D-1$.}
Since this is how (non-)supersymmetric AdS vacua are typically constructed in flux compactifications in string theory, see $e.g.$ \cite{VanRiet:2023pnx}, it is not a restrictive assumption.

The value of $F_2$ can in principle vary. Thus, we can have two different vacua supported by fluxes of the form 
\begin{equation}
\label{DF2}
F_2^{(\alpha)} = \alpha \epsilon_2\,, \qquad  F_2^{(\beta)} = \beta \epsilon_2 \,, 
\end{equation}
where $\epsilon_2$ is the Levi-Civita tensor for AdS$_2$ (we are suppressing spacetime indices). 
By invoking either the completeness hypothesis \cite{Polchinski:2003bq}, or the assumption that quantum gravity is unique, it follows that there should exist a domain wall particle or 0-brane interpolating between any two such vacua. Indeed, the vacuum expectation value of $F_2$ is constant, but it does not correspond to some propagating scalar field inside the 2D effective theory. This is because $F_2$ is a top form and, as such, it can change only by means of an interpolating domain wall. In this respect, these assumptions imply that pure AdS$_2$ (super)gravity should not exist if not supplemented by a domain wall.

This domain wall is characterized by a tension $T$ and a charge $Q$, but also by certain additional properties important for us. First, it is fundamental, which means that it cannot be completed into a smooth solitonic object within the 2D effective theory by including finitely-many new degrees of freedom. This implies that its tension defines an upper bound for the UV cutoff, $\Lambda_{UV}$, of the 2D effective theory,
\begin{equation}
\label{BPS1}
T>\Lambda_{UV}  \,. 
\end{equation}
Second, we assume that the domain wall interpolates between two supersymmetric vacua and thus it is BPS, $i.e.$~half of the bulk supersymmetry is preserved on its the world line, while the rest is non-linearly realized. As a consequence, its tension matches its charge,
\begin{equation}
\label{BPS2}
T = Q \,. 
\end{equation}
This requirement comes from preservation of supersymmetry on the world line. 
In higher dimensional theories this is equivalent to imposing that the gravitational attraction between two such objects cancels against the electric repulsion. 
This is arguably the most restrictive assumption for us, and it implies that our argument works only for supersymmetric AdS$_2$ vacua with at least two preserved supercharges. On the other hand, non-supersymmetric AdS vacua are believed to be unstable in quantum gravity \cite{Ooguri:2016pdq}.

At this point, let us notice that by combining the relations \eqref{BPS1} and \eqref{BPS2}, we obtain that 
\begin{equation}
\label{WGC}
Q > \Lambda_{UV}  \,. 
\end{equation}
This is the so-called magnetic weak gravity conjecture \cite{Arkani-Hamed:2006emk}. While its electric version may not hold in 2D, we find that something similar to the magnetic statement still survives. Indeed, it is equivalent to the requirement that the spectrum contains fundamental BPS objects.

Resuming our discussion, the fact that the domain wall interpolates between two AdS vacua means that the flux jump across it is
\begin{equation}
\Delta F_2 = Q \epsilon_2 \, , \qquad \text{with} \qquad  Q = \beta-\alpha  \, .
\end{equation}
If we are to respect flux quantization, this implies that 
\begin{equation}
\alpha = N \, Q \, , \qquad \beta = (N+1) \, Q \,,
\end{equation}
for some integer $N$. 
Then, from \eqref{AdSL} together with \eqref{BPS1} and \eqref{BPS2} we have that the AdS radius is bounded by 
\begin{equation}
\label{eq:LambdaLbound}
\frac{1}{L} = \frac12 N Q = \frac12 N T \gtrsim \Lambda_{UV} \, . 
\end{equation}
This relation is central to our argument.\footnote{For a macroscopic theory one would expect $N$ to be quite larger than ${\cal O}(1)$, which makes the BPS particle tension (and thus the cutoff) not only smaller, but considerably smaller than the AdS energy scale. We will see that indeed this happens in an explicit example in flux compactifications.} It states that the AdS length scale is bounded from above by the cutoff length and therefore the system is not a genuine 2D effective theory, for it probes length scales below the cutoff. 
This does not mean that the setup is problematic \emph{per se}, since it is protected by supersymmetry, but it does mean that it is not valid as a 2D effective theory in the Wilsonian sense. Unless it happens to be a consistent truncation, it must be supplemented by new physics, such as modes propagating in additional spacetime dimensions. In this sense, the most natural interpretation of \eqref{eq:LambdaLbound} is that supersymmetric AdS$_2$ flux vacua are not scale-separated, but rather need the existence of extra dimensions to be consistently defined quantum-mechanically. Let us stress, however, that this is not the only possibility, and in fact our argument rules out AdS$_2$ without invoking any geometric scale besides $L$; $\Lambda_{UV}$ can be non-geometric.

To get to the result \eqref{eq:LambdaLbound}, it is crucial that $M_P$ does not enter \eqref{eq:RLAdS2}. Otherwise, one could argue that scale separation might still be possible. 
Indeed, in $D\geq 2$ dimensions we have an electric flux $F_D\simeq Q \epsilon_D$, whose potential $C_{D-1}$ couples to a $(D-2)$-brane domain wall. Then, the flux-supported cosmological constant is $-L^{-2} M_P^{D-2} \simeq |F_D|^2$. Repeating our argument, with the assumption that $T>\Lambda_{UV}^{D-1}$, we are led to $L^{-2}M_P^{D-2} \simeq Q^2 \simeq T^2 M_P^{2-D}   > \frac{\Lambda_{UV}^{2D-2}}{M_P^{D-2}}$, which gives
\begin{equation}
  (M_P/\Lambda_{UV})^{D-2} > \Lambda_{UV} L\, .
\end{equation}
If $D>2$, this is not immediately incompatible with \eqref{eq:defscalesep}, for there can exist values of $L$ and $\Lambda_{UV}$ such that the right-hand side is greater than one. For $D=2$, the above relation is in conflict with the definition of scale separation \eqref{eq:defscalesep}.

Notice that the way that scale separation is excluded by our argument is similar, but not exactly the same,  as the way that it is excluded in 4D and 5D extended supergravity via the magnetic weak gravity conjecture \cite{Cribiori:2022trc,Cribiori:2023ihv}. 
Indeed, in this work we did not need to invoke the weak gravity conjecture at all.\footnote{Interestingly, analogs of black holes exist in 2D \cite{Brown:1986nm}.}  
Instead, it just appears in some form as a by-product of our assumptions, such as the requirement that there exist fundamental BPS domain walls in the spectrum.

Let us discuss further some of our main assumptions to better understand how restrictive these are. 
We start from the requirement that the cosmological constant be entirely supported by a top-form gauge field strength. Consider some constant vacuum energy $-2/L^2$. Then, to vary it one needs to trade it for a non-propagating field that gets non-trivial background on-shell values. 
The only field with such a property is a top-form gauge field strength. 
However, the requirement that we can have access to different values of the AdS radius does not automatically mean that the full cosmological constant is flux-supported, but only that at least a part of it is. 
Therefore, we are really dealing with a situation such as 
\begin{equation}
- \frac{2}{L^2} = \frac{1}{2} | F_2|^2 \pm  \lambda^2 \, ,
\end{equation}
where $ \lambda^2$ is the contribution to the cosmological constant which is not captured by the top-form gauge field strength. 
We discuss the two signs of the term $ \lambda^2$ independently. 
The positive sign, which reflects a positive contribution to the vacuum energy, is automatically beyond the domain of applicability of our argument because we are working on a supersymmetric AdS background and the only way to have a positive contribution to the cosmological constant is if there is at least one sector that spontaneously breaks supersymmetry.\footnote{One could ask what would happen if some contribution to the vacuum breaks supersymmetry partially. Since part of supersymmetry is preserved, then the same conditions apply as for unbroken supersymmetry. This happens because if one has a positive contribution to the cosmological constant then all supersymmetries are broken.}\footnote{If one works with  dS$_2$ supergravity \cite{Anninos:2023exn}, then even if such a theory can have positive contributions to the vacuum energy that preserve supersymmetry, it would be instead the negative contributions that break it. 
In other words, the dS$_2$ supergravity is also beyond the domain of applicability of our argument because it cannot give supersymmetric AdS vacua.} 
Next, we turn to the negative sign. A negative contribution to the cosmological constant can only make things worse as far as scale separation is concerned. Indeed, we have 
\begin{equation}
\frac{2}{L^2} = - \frac{1}{2} |F_2|^2 +  \lambda^2 \geq -\frac{1}{2} |F_2|^2 \gtrsim \Lambda_{UV}^2 \,. 
\end{equation}
Hence, we conclude that even if the vacuum energy has a contribution that is not flux-induced, the presence of a flux-induced term is enough to exclude scale separation for supersymmetric AdS vacua.

Next, we clarify what we mean when referring to the absence of scale separation. 
As we have seen, we postulate (or assume) the presence of a BPS domain wall particle in the AdS background. 
This object is light in the sense that its mass is of the order of the AdS scale. 
This tells us that the cutoff is of the order of the AdS scale, as we have argued.  
However, we can see that scale separation is also obstructed in the standard sense. 
Indeed, if such a domain wall particle exists and is BPS, one can have a bound state of an arbitrary number of them with mass 
\begin{equation}
m \sim \frac{n}{L} \,. 
\end{equation}
This is the characteristic mass spectrum of the Kaluza-Klein modes for a circle compactification with radius $L$. 
Hence, the 2D supergravity theory will effectively behave as if there is at least an additional $S^1$ with length of the order of the AdS$_2$ size.

Lastly, we would like to verify that our result is frame-independent, since otherwise we are not working with the correct energy scales. 
Actually this verification serves as a cross-check for the correct identification of the tension of the BPS domain wall particle. 
We identify the tension by looking at how the particle backreacts on the 2D AdS curvature; this is the strategy employed for example in the article of Brown and Teitelboim \cite{Brown:1988kg}. 
Frame-independence is not a generic property when comparing scales in a gravitational theory, neither should it be. For example, in 4D there is a well-defined frame, the Einstein frame, where the graviton is canonically normalized, and one can use it as the designated frame for the comparison of energy scales. 
One should be worried about the issue of frame-(in)dependence because our 2D theory (as any 2D theory)  does not have a well-defined Einstein frame and hence no designated frame in which we should measure scales and distances. 
This means that we can at best compare energy scales and if the result is physical, it should be frame-independent.

We recall that the Einstein equation that links the AdS scale $L$ and the domain wall tension $T$ is 
\begin{equation}
R = - \frac{2}{L^2} + \frac{T}{\sqrt{-g}}\int_{\mathcal{W}_1}  d \tau \, \delta^{(2)} (x - X(\tau))  \sqrt{-g_{\mu \nu} \dot X^\mu(\tau) \dot X^\nu (\tau)} \,, 
\end{equation}
up to unimportant coefficients and where $\mathcal{W}_1$ is the particle worldline. Compared to \cite{Brown:1988kg} we absorbed the gravitational coupling into $T$ since that is the actual physical scale that characterizes the backreaction of the domain wall on the metric.\footnote{For example, if in 4D we have $M_P^2 R \sim c^4$ for some constant $c$ (with $[c]=1$), the actual physical scale as far as gravity is concerned is $c^2/M_P$, because that is the scale backreacting on the metric and eventually characterizing the (A)dS scale. 
Indeed, one usually writes $c^4 = 3 M_P^2 H^2$, where $H$ is the physical scale.} 
If we perform a rescaling of the form 
\begin{equation}
g_{\mu\nu} \to  g_{\mu\nu} w^2 \, ,
\end{equation}
we find then in terms of the new metric 
\begin{equation}
R = - 2 \frac{w^2}{L^2} + \frac{w \,T}{\sqrt{-g}} \int_{\mathcal{W}_1}  d \tau \, \delta^{(2)} (x - X(\tau)) \sqrt{-g_{\mu \nu} \dot X^\mu (\tau) \dot X^\nu (\tau) } \,. 
\end{equation}
This means that we can identify the following frame-dependent energy scales: 
\begin{equation}
L' =  L/w \,, \qquad T'= w T \,,
\end{equation}
from which we see that crucially 
\begin{equation}
T' L' =  T L  \, 
\end{equation}
and therefore our result is frame-independent. 
In other words, if $1/L \sim T$, then it holds that $1/L'\sim T'$. 
This means that the AdS energy is comparable to the domain wall tension independently of which frame is adopted.

\section{2D supergravity analysis}

In the previous section, we presented our general argument in a schematic manner. We are now going to implement it in concrete 2D supergravity models taking care of the subtleties that arise. We will focus on $\mathcal{N}=(1,1)$ supergravity in 2D, namely two supercharges, but our results can be directly extended to setups with more supersymmetry.

\subsection{JT supergravity and one-form dilaton multiplet}

The $\mathcal{N}=(1,1)$ supersymmetric extension of Jackiw-Teitelboim (JT) gravity \cite{Teitelboim:1983ux,Jackiw:1984je} contains the gravity multiplet, consisting of  a zweibein $e_{\mu}^a\,,$ a Majorana gravitino $\psi_{\mu}$ and an auxiliary real scalar $A$ \cite{Howe:1978ia}, and a dilaton muliplet, containing the dilaton $\phi$, a Majorana fermion $\lambda$ (dilatino) and an auxiliary real scalar $F$. The off-shell component form action is  \cite{Chamseddine:1991fg,Astorino:2002bj} 
\begin{equation}
\begin{aligned}
    \label{JTSGm}
    S = \int d^2 x \, e &  \left( \phi R  - 2 \left( A + \frac{1}{L} \right) F - \frac{2}{L} \phi A + \frac{1}{2L} \phi \epsilon^{\mu \nu}  \bar{\psi}_{\mu} \gamma_3 \psi_{\nu}  \right. \\ & \quad \left.- 2 \epsilon^{\mu \nu} \bar{\lambda} \gamma_3 D_{\mu} \psi_{\nu} + \frac{1}{L} \bar{\lambda} \gamma^{\mu} \psi_{\mu} \right) , 
\end{aligned}
\end{equation}
and we refer to appendix \ref{eq:SUGRAconv} for more information about the notation.  
Varying the action \eqref{JTSGm} with respect to the fields of the dilaton multiplet we obtain
\begin{align}
    \label{Feom1}
    \delta F & : \ \ A = - \frac{1}{L} \,, \\ 
    \label{lameom1}
    \delta \lambda & : \ \ \epsilon^{\mu \nu} \hat{D}_{\mu} \psi_{\nu} = 0 \, , \\
    \label{phieom1}
    \delta \phi & : \ \ R = - \frac{2}{L^2} - \frac{1}{2L} \epsilon^{\mu \nu} \bar{\psi}_{\mu} \gamma_3 \psi_{\nu} \, .
    \end{align}
The last of these equations indicates the presence of an AdS$_2$ vacuum with radius $L$.

To give the cosmological constant a dynamical role while preserving supersymmetry, we can dualize an auxiliary field to a two form. In appendix \ref{app:1formSUGRA} we show that doing this operation on the auxiliary field $A$ of the gravity multiplet does not give the desired effect. Hence, we dualize below the real auxiliary scalar of the dilaton multiplet, $F$, to a two-form $H_2$ that is locally  $H_2 = d B_1$, where $B_1$ is a one-form, since any two-form is exact in 2D.  More precisely, what is to be dualized is a specific combination of $F$ and $A$, together with fermionic terms, namely 
\begin{equation}
    \label{starH2main}
    \star H_2 = \frac{1}{2} \epsilon^{\mu \nu} H_{\mu \nu} = \epsilon^{\mu \nu} \partial_{\mu} B_{\nu} = F + \phi A - \frac{1}{2} \bar{\lambda} \gamma^{\mu} \psi_{\mu} - \frac{1}{4} \phi \epsilon^{\mu \nu} \bar{\psi}_{\mu} \gamma_3 \psi_{\nu} \,,
\end{equation}
or, equivalently,
\begin{equation}
    \label{Hmunumain}
    H_{\mu \nu}  = - \left( F + \phi A - \frac{1}{2} \bar{\lambda} \gamma^{\rho} \psi_{\rho} - \frac{1}{4} \phi \epsilon^{\rho \sigma} \bar{\psi}_{\rho} \gamma_3 \psi_{\sigma} \right) \epsilon_{\mu \nu}\,.
\end{equation}
This combination is the field-dependent coefficient of $-2/L$ in the action \ref{JTSGm}.
Taking the supersymmetry variation of this expression leads to an integrable equation that implies the following local supersymmetry transformation rule for the one-form $B_1$:
\begin{equation}
\label{dBmu}
    \delta_{\epsilon} B_{\mu} = \frac{1}{2} \bar{\epsilon} \gamma_{\mu} \gamma_3 \lambda - \frac{1}{2} \phi \bar{\epsilon} \gamma_3 \psi_{\mu} \,.
 \end{equation}
The commutator of two local supersymmetry transformations acting on $B_{\mu}$ is then
\begin{equation}
    \label{QQBcomm}
    [\delta_Q (\epsilon_1),\delta_Q (\epsilon_2) ] B_{\mu} = \delta_{\text{gct}} (\xi^{\nu}) B_{\mu} - \delta_{\text{gauge}} (\xi^{\nu} B_{\nu}) B_{\mu} - \delta_Q (\xi^{\nu} \psi_{\nu}) B_{\mu} + \delta_{\text{gauge}} \left( \frac{1}{2} \phi \bar{\epsilon}_1 \gamma_3 \epsilon_2 \right) B_{\mu} \,,
\end{equation}
where $\delta_{\text{gauge}}$ is an abelian gauge transformation that acts on $B_{\mu}$ as $\delta_{\text{gauge}}(\Lambda) B_{\mu} = \partial_{\mu} \Lambda$.

Substituting \eqref{starH2main} into $\eqref{JTSGm}$ we obtain the JT supergravity action supplemented with the one-form,
\begin{equation}
\begin{aligned}
\label{S1formdilmain}
    S_{\text{1-form}} = \int d^2 x \, e \,  &  \bigg( \phi R - 2 A \star H_2 + 2 \phi A^2  - A \bar{\lambda} \gamma^{\mu} \psi_{\mu}   \\ &-\left. \frac{1}{2} \phi A \epsilon^{\mu \nu} \bar{\psi}_{\mu} \gamma_3 \psi_{\nu} - 2 \epsilon^{\mu \nu} \bar{\lambda} \gamma_3 D_{\mu} \psi_{\nu} \right) , 
\end{aligned}
\end{equation}
where we have omitted a total spacetime derivative proportional to $e \star H_2 $. This term is crucial when substituting the solution of the equation of motions back into the action, but since we are not going to perform this step, we omit it in what follows. 
Varying the action \eqref{S1formdilmain} with respect to $A$ we find 
\begin{equation}
    \label{AEL}
    A = \frac{1}{2 \phi} \star H_2  \, 
\end{equation}
up to terms that involve the dilatino and the gravitino. Then, the bosonic sector of the Euler-Lagrange equation for $\phi$ is
\begin{equation}
    \label{dilEL}
    R = - \frac{1}{2 \phi^2} (\star H_2)^2 \,,
\end{equation}
while the equation of motion for the one-form $B_1$ reads
\begin{equation}
    \label{BEL}
    \partial_{\mu} \left( \frac{1}{\phi} \star H_2 \right) = 0 ,
\end{equation}
up to terms with fermionic fields. From the last equation it follows that 
\begin{equation}
    \label{starH2phi-1=cst}
    \frac{1}{\phi} \star H_2 = c \,,
\end{equation}
where $c$ is a real constant of integration. Then, equation \eqref{dilEL} implies that
\begin{equation}
    \label{AdS2}
    R = -\frac{1}{2} c^2 \,,
\end{equation}
which for $c\neq0$ describes an $\rm{AdS}_2$ spacetime with length scale $L$ related to the constant value of $\frac{1}{\phi} \star H_2$ by 
\begin{equation}
    \label{c=c(L)}
    c^2 = \frac{4}{L^2} \,.
\end{equation}
To summarize, by performing the replacement \eqref{Hmunumain} into the original $\mathcal{N}=(1,1)$ JT supergravity action, we constructed a novel supersymmetric model governed by \eqref{S1formdilmain}. Since we replaced an auxiliary field, we did not change the spectrum of dynamical degrees of freedom and the novel action ought to be on-shell equivalent to the one we started from. Crucially, the model \eqref{S1formdilmain} admits AdS$_2$ vacua in which the radius $L$ is set to a parameter $c$ that arises dynamically by integrating out the one-form $B_1$. In this sense, the cosmological constant has become dynamical. 

\subsection{Coupling to a BPS domain wall particle}

The one-form $B_1$ can be coupled to a domain wall particle whose worldline divides the 2D target spacetime into two regions with different cosmological constants. The $\kappa$-symmetric action that describes the coupling of the one-form dilaton multiplet to a BPS domain wall in curved 2D $\mathcal{N}=(1,1)$ superspace can be found in appendix \ref{particledil}. In this subsection, we only give its bosonic sector, which reads
\begin{equation}
    \label{Sspboson}
    S_{\text{sp,bos}} = - |Q| \int_{{\cal W}_1} d \tau \sqrt{- g_{\mu \nu} (X^{\rho}) \dot{X}^{\mu} \dot{X}^{\nu}} |\phi(X^{\rho})| + Q \int_{{\cal W}_1} d \tau B_{\mu} (X^{\nu}) \dot{X}^{\mu} \,,
\end{equation}
where ${\cal W}_1$ denotes the worldline of the particle, which is parametrized by the coordinate $\tau$ and whose embedding into the 2D background spacetime is described by the parametric equations $x^{\mu} = X^{\mu} (\tau)$.
Also, $\dot{X}^{\mu} \equiv \frac{d X^{\mu}}{d\tau}$, while $Q$ is the real charge of the particle under the one-form $B_1$, which sets the tension $T= |Q|$. 

We then consider the action that is the sum of \eqref{S1formdilmain} and \eqref{Sspboson}, and specify the associated equations of motion with vanishing fermionic fields. First, the Euler-Lagrange equation for the auxiliary scalar field of the supergravity multiplet is still \eqref{AEL}. Then, varying the action with respect to the dilaton $\phi$ we obtain
\begin{equation}
       \label{dilELsource}
    R = - \frac{1}{2 \phi^2} (\star H_2)^2 + e^{-1} |Q| \frac{\phi}{|\phi|} \int_{{\cal W}_1} d \tau \, \delta^{(2)} (x - X(\tau)) \sqrt{-g_{\mu \nu} \dot{X}^{\mu}(\tau) \dot{X}^{\nu}(\tau)} \,,
\end{equation}
while the equations of motion for the one-form $B_1$ read
\begin{equation}
    \label{BELsource}
    \partial_{\mu} \left( \frac{1}{\phi} \star H_2 \right) = Q e^{-1} \epsilon_{\mu \nu} \int_{{\cal W}_1} d \tau \, \delta^{(2)} (x - X(\tau)) \dot{X}^{\nu}(\tau) .
\end{equation}
Varying the action with respect to the metric, $g_{\mu \nu}$, we find 
\begin{equation}
    \label{metricEL}
    \nabla^{\mu} \partial^{\nu} \phi - g^{\mu \nu} \nabla_{\rho} \partial^{\rho} \phi  + \frac{1}{4 \phi}  (\star H_2)^2 g^{\mu \nu} + \frac{1}{2} e^{-1} |Q \phi| \int_{{\cal W}_1} d \tau \, \delta^{(2)} (x - X(\tau))  \frac{\dot{X}^{\mu} \dot{X}^{\nu}}{\sqrt{- g_{\rho \sigma} \dot{X}^{\rho} \dot{X}^{\sigma}}}  = 0 \, .
        \end{equation}
Finally, the Euler-Lagrange equations for $X^{\mu}$ are 
\begin{align}
\label{XEL}
\frac{d}{d \tau} \left( \frac{|\phi(X)|}{\sqrt{- g_{\rho \sigma} (X)\dot{X}^{\rho} \dot{X}^{\sigma}}}  g_{\mu \nu} (X) \dot{X}^{\nu} \right) = & \,\frac{1}{2} \frac{|\phi(X)|}{\sqrt{- g_{\lambda \sigma} (X) \dot{X}^{\lambda} \dot{X}^{\sigma}}} \dot{X}^{\nu} \dot{X}^{\rho} \partial_{\mu} g_{\nu \rho} (X) \nonumber \\
& - \sqrt{- g_{\nu \rho} (X)\dot{X}^{\nu} \dot{X}^{\rho}} \frac{\phi(X)}{|\phi(X)|} \partial_{\mu} \phi (X) \\ &  + \frac{Q}{|Q|} H_{\mu \nu} (X) \dot{X}^{\nu} \,. \nonumber
\end{align}
In particular, equation \eqref{BEL} implies that away from the particle, where $x^{\mu} \neq X^{\mu} (\tau)$, the quantity $\phi^{-1} \star H_2$ is constant, since the delta function on its right-hand side vanishes.  The presence of this delta function means that the value of $\phi^{-1} \star{H}_2$ is different in the two regions into which the 2D background spacetime is divided by the worldline of the particle. Let $x^{\mu} = (t,x)$ be the coordinates of the 2D target spacetime and $X^{\mu} = (\mathcal{T},\mathcal{X})$. If the particle lies in the axis $x=0$, that is $\mathcal{X}=0$, then in the static gauge, in which $\mathcal{T} = \tau$, the $x$ component of \eqref{BELsource} becomes
\begin{equation}
    \label{BELstatic}
    \partial_x (\phi^{-1} \star H_2) = - Q \delta(x) \,.
\end{equation}
Let $\alpha_{+}$ and $\alpha_{-}$ be the constant values of $\phi^{-1} \star H_2$ in the regions $x>0$ and $x<0$ respectively. Integrating the differential equation \eqref{BELstatic} over an interval $[-\epsilon,\epsilon]$, where $\epsilon$ is a positive real number, we find 
\begin{equation}
    \label{fluxdiscont}
    \alpha_{+} - \alpha_{-} = - Q \,.
\end{equation}
If neither of $\alpha_{\pm}$ is zero, then the field equation \eqref{dilELsource} implies that the regions $x>0$ and $x<0$ are both $\rm{AdS}_2$ spaces with length scales $L_{+}$ and $L_{-}$ respectively, which are given by 
\begin{equation}
    \label{Lpm}
    \frac{1}{L_{\pm}}  = \frac{1}{2} |\alpha_{\pm}| \,.
\end{equation}
From equations \eqref{fluxdiscont} and \eqref{Lpm} it follows that 
\begin{equation}
    \label{noscalesep}
    \max \left\{\frac{1}{L_{+}},\frac{1}{L_{-}} \right\} \geq \frac{1}{2} \left( \frac{1}{L_{+}} + \frac{1}{L_{-}}\right) = \frac{1}{4} (|\alpha_{+}|+|\alpha_{-}|) \geq \frac{1}{4} |\alpha_{+} - \alpha_{-}| = \frac{1}{4} |Q| > \frac{1}{4} \Lambda_{\rm{UV}}.
   \end{equation}
Therefore, the cosmological constant in at least one of the regions $x>0$, $x<0$ is of the order of or greater than the ultraviolet cutoff, which means the absence of scale separation, as explained in section \ref{sec:arg}.

Here, we have used $T=|Q|$ as the tension of the domain wall particle because it is the coefficient that appears in the 2D Einstein equations. 
In addition, we are treating this scale as a cutoff because otherwise the vibration modes of the particle will enter the effective description (see $e.g.$ \cite{Lanza:2020qmt} for a similar discussion related to fundamental strings and membranes in four dimensions). 
One can also think of this cutoff entering due to the non-linear supersymmetry on the worldline, which is present since the domain wall is BPS.

\subsection{BPS domain wall solutions}
\label{domainwall}

We would like to solve  the equations of motion \eqref{dilELsource}-\eqref{XEL} for a charged domain wall particle that lies in the axis $x=0$, which means that ${\cal X}=0$. 
We perform this analysis to verify that the domain walls exist and that they are indeed BPS as we have claimed. 
Time translation symmetry implies that the spacetime metric must be of the form  
\begin{equation}
    \label{ds2ans}
    ds^2 = - e^{2 f(x)} d t^2 + dx^2 , 
\end{equation}
while the one-form $B_1$ and the dilaton $\phi$ have to be such that (we choose the static gauge $\mathcal{T}=\tau$)
\begin{equation}
    \label{Bphians}
    B_1 = B(x) dt \qquad \text{and} \qquad \phi = \phi(x) \,.
\end{equation}

The vielbein one-forms $e^a = e^a_{\mu} dx^{\mu}$ that correspond to the metric \eqref{ds2ans} are $e^0 = e^f dt $ and $e^1 = dx$, so the components of the relevant torsion-free spin connection, $\omega_{\mu} (e) =e_{\mu}^a \epsilon^{\nu\rho} \partial_{\nu} e_{a \rho} \,$, are given by $\omega_t (e) = - e^f f'$ and $\omega_x (e) =0 $, where a prime denotes a derivative with respect to $x$. 
The Ricci scalar associated to the metric \eqref{ds2ans} is
\begin{equation}
    \label{Ricciscans}
    R = - 2 \left[ f^{''} + (f')^2 \right]  ,
\end{equation}
the ansatz for the spacetime metric and the one-form imply that 
\begin{align}
    \label{fluxans}
    \star H_2 = e^{-f} B',
\end{align}
and thus the Euler-Langrange equation for the dilaton, \eqref{dilELsource}, becomes
\begin{equation}
    \label{dilELans}
    f'' + (f')^2 - \frac{1}{4} (\phi^{-1} e^{-f} B')^2 = - \frac{1}{2} |Q| \frac{\phi(0)}{|\phi(0)|} \delta (x)\,.
\end{equation}
The $t$-component of the equation of motion for the one-form $B_1$ is trivially satisfied under the above assumptions, while its $x$-component gives
\begin{equation}
    \label{BELxcomp}
    (\phi^{-1} e^{-f} B')^{'} = - Q \delta (x) \,.
\end{equation}
The $tt$- and $xx$-components of the metric field equation reduce to 
\begin{equation}
    \label{Einsttans}
    \phi^{-1} \phi^{''} - \frac{1}{4} (\phi^{-1} e^{-f} B')^2 = - \frac{1}{2} |Q| \frac{|\phi(0)|}{\phi(0)} \delta(x) 
\end{equation}
and 
\begin{equation}
    \label{Einsxxans}
      \phi^{-1} f' \phi' - \frac{1}{4} (\phi^{-1} e^{-f} B')^2 = 0  
\end{equation}
respectively, while the $tx$-component holds trivially.

In appendix \ref{SolutionofODEs}, we show that if $Q \phi (0) >0$, the ordinary differential equations \eqref{dilELans}-\eqref{Einsxxans} are solved by 
\begin{align}
\label{phiQphi>0main}
\phi(x) &= 
\phi (0) \exp \left[ \frac{1}{2} (\alpha + Q H(-x)) x \right] , \\
\label{fQphi>0main}
f(x) &= 
    \ln |\phi (x)| \,, \\
    \label{BQphi>0main}
B (x) & =  
(\sgn Q )\phi^2 (0) \exp [(\alpha + Q H(-x))x] \,, 
    \end{align}
where $\alpha$ is an arbitrary real constant, $H$ is the Heaviside step function (with $H(0)=1$) and we have required the functions $\phi$, $f$ and $B$ to be continuous at $x=0$. On the other hand, if $Q \phi(0) <0$, a general solution to equations \eqref{dilELans}-\eqref{Einsxxans} that is continuous at $x=0$ is 
\begin{align}
\label{phiQphi<0main}
\phi(x) &=  
\phi (0) \exp \left[- \frac{1}{2} (\alpha + Q H(-x)) x \right] , \\
\label{fQphi<0main}
f(x) &= 
    \ln |\phi (x)| \,, \\
    \label{BQphi<0main}
B (x) & =   
(\sgn Q )\phi^2 (0) \exp [-(\alpha + Q H(-x))x] \,.
    \end{align}
Provided that $\alpha \neq 0, -Q$, the metric \eqref{ds2ans} with $f$ given by equation \eqref{fQphi>0main} or \eqref{fQphi<0main} describes two $\rm{AdS}_{2}$ domains with length scales $2|\alpha|^{-1}$ and $2|\alpha+Q|^{-1}$, which are the regions $x>0$ and $x<0$ respectively. In fact, this metric is written in horospherical coordinates \cite{Duff:1994fg}.

    Our next task is to determine the Killing spinors associated with the purely bosonic background described by equations \eqref{ds2ans}, \eqref{Bphians} and \eqref{phiQphi>0main}-\eqref{BQphi>0main} or \eqref{phiQphi<0main}-\eqref{BQphi<0main}. To this purpose, we set the local supersymmetry transformations of the fermionic fields $\lambda$ (dilatino) and $\psi_{\mu}$ to zero. Using the supersymmetry transformation rules \eqref{dpsimu} and \eqref{delambda} and equations \eqref{starH2main} and \eqref{AEL} we find that the local supersymmetry variations of the fermions $\lambda$ and $\psi_{\mu}$ in a backgound where these fields vanish are 
    \begin{align}
        \label{dlf=0}
        \delta_{\epsilon} \lambda &= \frac{1}{2} \gamma^{\mu} \epsilon \partial_{\mu} \phi + \frac{1}{4} \star H_2 \, \epsilon \,, \\
        \label{dpsif=0}
        \delta_{\epsilon} \psi_{\mu} &= \partial_{\mu} \epsilon + \frac{1}{2} \omega_{\mu} (e) \gamma_3 \epsilon + \frac{1}{4 \phi} \star H_2 \, \gamma_{\mu} \epsilon \,.
    \end{align}
    Notice that we used the fact that on-shell and up to fermionic terms which vanish on a supersymmetric vacuum, we have $F=*H_2 -\phi A = \frac12 *H_2$.

We first examine the case in which $Q \phi(0) >0\,$. Using the ansatzes \eqref{ds2ans} and \eqref{Bphians} for the metric and one-form respectively, and the solution \eqref{phiQphi>0main}-\eqref{BQphi>0main} to the field equations, we deduce that \eqref{dlf=0} vanishes if and only if 
\begin{equation}
    \label{BPS-g1}
    \epsilon = - \gamma_1 \epsilon \,,
\end{equation}
which means that the Majorana spinor parameter $\epsilon$ has only one independent component. Therefore, the particle preserves half of the $\mathcal{N}=(1,1)$ target spacetime supersymmetry, so it is a BPS solution.

We also require that the local supersymmetry transformation  \eqref{dpsif=0} of the gravitino vanish,
\begin{equation}
    \label{dpsi=0}
    \delta_{\epsilon} \psi_{\mu} = 0 \, .
\end{equation}
Given the projection condition \eqref{BPS-g1}, the $t$-component of equation \eqref{dpsi=0} implies that 
\begin{equation}
    \label{dte=0}
    \partial_t \epsilon  = 0 \,,
\end{equation}
that is $\epsilon$ depends only on the spatial coordinate $x$. On the other hand, the $x$ component of  \eqref{dpsi=0} gives 
\begin{equation}
    \label{dxepsilon}
    \epsilon' - \frac{1}{4} (\alpha + Q H(-x)) \epsilon =0 \,,
\end{equation}
whose general solution is 
\begin{equation}
\label{Killingeps1}
    \epsilon (x) =  \exp\left[ \frac{1}{4} (\alpha + Q H(-x)) x \right]\eta \,,
\end{equation}
where $\eta$ is an arbitrary constant Majorana spinor satisfying the condition 
\begin{equation}
    \label{eta=-g1eta}
    \eta = - \gamma_1 \eta \,,
\end{equation}
which follows from \eqref{BPS-g1}. Equations \eqref{Killingeps1} and \eqref{eta=-g1eta} describe the Killing spinors of the $\rm{AdS}_2$ background spacetime with a particle lying in the axis $x=0$, if the charge $Q$ has the same sign as the value of the dilaton at $x=0$. 

On the other hand, if $Q \phi (0) < 0$, the corresponding $\text{AdS}_2$ background is described by equations \eqref{ds2ans}, \eqref{Bphians} and \eqref{phiQphi<0main} - \eqref{BQphi<0main}. In this bosonic background, the supersymmetry variation \eqref{dlf=0} of the dilatino vanishes if and only if 
\begin{equation}
    \label{BPS+g1}
    \epsilon = \gamma_1 \epsilon \,,
\end{equation}
so the particle solution preserves again half of the $\mathcal{N}=(1,1)$ bulk supersymmetry. Then, the $t$-component of the Killing spinor equation \eqref{dpsi=0} implies that the Majorana spinor parameter $\epsilon$ is independent of the time coordinate $t$ and the $x$-component of \eqref{dpsi=0} becomes 
\begin{equation}
\label{dxe'}
    \epsilon' + \frac{1}{4} (\alpha + Q H(-x)) \epsilon =0 \,,
    \end{equation}
which is solved by 
\begin{equation}
    \label{Killsp2}
     \epsilon (x) =  \exp\left[- \frac{1}{4} (\alpha + Q H(-x)) x \right]\eta \,,
\end{equation}
where $\eta$ is an arbitrary constant Majorana spinor that obeys the projection condition 
\begin{equation}
    \label{g1eta=eta}
    \eta = \gamma_1 \eta \,,
\end{equation}
by virtue of \eqref{BPS+g1}.

\section{Examples from flux compactifications}

In this section, we show how our general argument forbidding scale-separated, supersymmetric AdS$_2$ flux vacua is realized in compactifications of string theory. We focus on the type II superstring, since this is the most studied for what concerns scale separation, but other duality frames can be considered as well.

\subsection{An example in type IIA} 

Several explicit string theory examples in which our argument is realized can be found in \cite{Lust:2020npd}. For concreteness, let us discuss the setup with a supersymmetric solution of the form
\begin{equation}
\label{eq:IIAexmaple}
\rm{AdS}_2 \times \rm{S}^2 \times \rm{T}^6 \,. 
\end{equation}
The observation of \cite{Lust:2020npd} is that the AdS$_2$ and the S$^2$ length scales cannot be disentangled, so there is no complete scale separation between the compact and the non-compact spacetime directions. 
In our analysis below, we will not really invoke the fact that the Kaluza-Klein modes of S$^2$ are light compared to the AdS$_2$ scale. Instead, we will show that it is already the AdS$_2$ itself that hinders scale separation. 
This is in accordance with the bottom-up approach we presented in the previous sections.

The setup consists of type IIA supergravity with RR fields $F_{p}$, but no (localized) sources. Indeed, the geometry \eqref{eq:IIAexmaple} arises as the near-horizon limit of a system of $D0/D4$-branes wrapped along $T^6$. 
In the near-horizon limit, branes dissolve into flux and one has electric $F_2$ and $F_6$ fluxes, $i.e.$ filling the AdS$_2$ directions, while $F_0=0$. Alternatively, one can dualize either of them into the magnetic fluxes $F_8$ and $F_4$ respectively, with all their legs along the compact space. Indeed, we will right away work with $F_2$ and $F_4$ turned on. Besides the metric \eqref{eq:IIAexmaple}, the other NS fields are
\begin{equation}
H_3=0 \, , \qquad \varphi = \varphi_0 \,,
\end{equation}
with $\varphi_0$ a constant 10D dilaton. The equations of motion in the Einstein frame without any sources and for our given flux choices take the form\footnote{We are following the conventions of \cite{Danielsson:2009ff,Blaback:2010sj}. The coordinates of the 10D spacetime are labeled by $M,N,\dots \,=0,1,\dots,9$, while for the coordinates of the 2D external and the 8D internal spaces we use indices $\mu,\nu,\dots \, = 0,1$ and $i,j,\dots \, = 2,\dots,9$. We also use the notation $|A_p|^2\equiv\frac{1}{p!} A_{M_1 \dots M_p} A^{M_1 \dots M_p}$ 
and $|A_p|_{MN}^2\equiv\frac{1}{(p-1)!} A_{M M_2 \dots M_p} A_N{}^{M_2 \dots M_p}$, such that $g^{MN}|A_p|^2_{MN} = p|A_p|^2$. } 
\begin{align}
\label{GR}
R_{MN} &= e^{3 \varphi/2} \left( \frac12 |F_2|^2_{MN} - \frac{1}{16} g_{MN} |F_2|^2 \right) 
+ e^{\varphi/2} \left( \frac12 |F_4|^2_{MN} - \frac{3}{16} g_{MN} |F_4|^2 \right), 
\\
\label{DIL}
0 &= \frac34 e^{3 \varphi/2} |F_2|^2 + \frac14 e^{\varphi/2} |F_4|^2 \,, 
\\ 
\label{RRBI}
0&=dF_p = d(\star F_p)   \,,  \qquad \text{with} \quad p=2,4 \,. 
\end{align}
Specifically, equation \eqref{GR} is the trace-reversed Einstein equation, 
equation \eqref{DIL} follows from the variation of the 10D action with respect to the dilaton $\varphi$, 
and equations \eqref{RRBI} are the Bianchi identities and equations of motion for the RR fields. 
The equation of motion for the Kalb-Ramond field, $B_2$, is automatically satisfied, since the magnetic $F_4$ flux always threads the $\rm{S}^2$, while the electric $F_2$ and $F_6$ are always spacetime-filling. 
Therefore, $F_2 \wedge F_6 =0$ and $F_4 \wedge F_4 =0$ because all the $F_4$ flux wraps 4-cycles of the form $\rm{S}^2 \times \rm{T}^2$, and we have only a single $\rm{S}^2$ and three possible 2-tori. 
Notice that the dilaton field equation \eqref{DIL} explicitly implies that, indeed, if the flux $F_2$ is electric, then $F_4$ has to be magnetic. 
For completeness, let us note that the relation between magnetic and electric fluxes is governed by the 10D Poincar\'{e} duality rule $e^{(5-n) \varphi /2}F_n = (-1)^{(n-1)(n-2)/2}\star F_{10-n}$.

From these equations we can see why the external space is AdS$_2$ and why the inverse of its length scale $L^{-1}$ is proportional to the modulus of the $F_2$ flux.  Once we contract the $\mu \nu$-component of equation \eqref{GR} with the inverse of the external spacetime metric, $g^{\mu \nu}$, we find
\begin{equation}
R_{2D} =g^{\mu  \nu} R_{\mu \nu} = \frac78 e^{3 \varphi/2} |F_2|^2 - \frac{3}8 e^{\varphi/2} |F_4|^2 \,, 
\end{equation}
which, given the equation of motion for the dilaton, implies that 
\begin{equation}
R_{2D} = 2e^{3\varphi /2}|F_2|^2 = - 2e^{-3\varphi /2}|F_8|^2.
\end{equation}
Hence, this AdS$_2$ vacuum is flux-supported, as required by our argument of section \ref{sec:arg}. 
For completeness note that for the dilaton we have 
\begin{equation}
e^{2 \varphi} = \frac{3 |F_8|^2}{|F_4|^2} \,. 
\end{equation}

The next step is to link the electric flux $F_2$, or the magnetic $F_8$, to the tension of some BPS domain wall. To this purpose, we look at the Bianchi identities, which also give the conditions on flux quantization. 
Since we are in 2D and the domain wall is a particle, the most natural BPS fundamental such object is a $D0$-brane. Its Bianchi identity is
\begin{equation}
d F_8 = \mu_{0} \delta_9 \,, 
\end{equation}
where $\mu_0$ is the charge and $\delta_9$ is a Dirac delta function nine-form with legs transverse to the worldline of the brane. 
The flux quantization condition reads 
\begin{equation}
\int_{X_8} F_8 = (2 \pi l_s)^7 f_8 \,, 
\end{equation}
where $X_8=\rm{S}^2 \times \rm{T}^6$ denotes the eight-dimensional internal manifold, $l_s = \sqrt{\alpha'}$ is the string length 
and $f_8 \in \mathbb{Z}$ is an integer number. 
This is indeed a rewriting of the flux in terms of the {localized} $D0$-brane charge. 
Thus, taking flux quantization into account and using $|F_8|^2 
= \frac{1}{8!} F_{i_1 \dots i_8} F_{j_1 \dots j_8} g^{i_1 j_1}\dots  g^{i_8 j_8} 
= \frac{\left( (2 \pi l_s)^7 f_8 \right)^2}{(\text{Vol}[X_8])^2 } $,
where $\text{Vol}[X_8]$ is the product of all of the radii in the internal space metric, we find 
\begin{equation}
\label{Rextqu}
R_{2D} = - 2 \Bigg{(} \frac{(2 \pi l_s)^7 f_8}{e^{3 \varphi/4} \text{Vol}[X_8]} \Bigg{)}^2 \,.
\end{equation} 
Let us set the 10D gravitational coupling constant equal to $2 \kappa_{10}^2 = (2 \pi)^7 l_s^8 = 1 $, so that $l_s = (2 \pi)^{-7/8} $ and the AdS$_2$ radius can be read off from \eqref{Rextqu} to be 
\begin{equation}
    \label{LAdS2}
    \frac{1}{L} = \frac{(2 \pi)^{7/8} |f_8|}{\text{Vol}[X_8]} e^{-3 \varphi/4}.
\end{equation} 
We can now show that this scale corresponds to the tension of a $D0$-brane smeared over the compact directions. The appearance of a smeared object is to be expected from a 2D point of view, for the $D0$-brane is localized in 10D and, from a bottom-up perspective, one has not really access to its position in the internal space. We leave for future work the study of how backreaction corrections allow one to go beyond the smeared approximation \cite{Junghans:2020acz}.

In general, one way to identify the tension of an extended object is by performing the direct dimensional reduction and studying the resulting effective action. 
However, in 2D this strategy is not feasible, for there is no kinetic term for gravity and hence no Einstein frame, as already mentioned. 
Instead, we can identify the tension by  deducing it from its backreaction on the 2D curvature. 
Let us consider a $D0$-brane in 10D with worldline $\mathcal{W}_1$ parametrized by a coordinate $\tau$. The embedding of $\mathcal{W}_1$ into 10D spacetime is described by parametric equations of the form $x^M= X^M(\tau)$. In the Einstein frame, the action for such a source  is
\begin{equation}
    \label{SD0}
    S_{\text{D0}} = - T_0 \int_{\mathcal{W}_1} d \tau e^{-\frac{3}{4} \varphi} \sqrt{-g_{MN} \dot{X}^M \dot{X}^N}  + \mu_0 \int_{\mathcal{W}_1} C_1 \,,
\end{equation}
where $T_0 = \frac{1}{l_s}=(2 \pi)^{7/8}$ is the tension (for $2 \kappa_{10}^2=1$), $\dot{X}^M \equiv \frac{dX^M}{d\tau}$ and $C_1$ is the RR one-form. 
The energy-momentum tensor associated with the D0-brane is 
\begin{equation}
\begin{aligned}
    \label{TD0MN}
    T^{\text{D}0}_{MN} &\equiv -\frac{2}{\sqrt{-g_{(10)}}} \frac{\delta S_{\text{D}0}}{\delta g^{MN}}\\
    &= \frac{1}{\sqrt{-g_{(10)}}} T_0\, g_{MP} \,g_{NQ} \,e^{-\frac{3}{4} \varphi} \int_{\mathcal{W}_1} d \tau \, \delta^{(10)} (x-X(\tau)) \frac{\dot{X}^P \dot{X}^Q}{\sqrt{-g_{RS} \dot{X}^R \dot{X}^S}}  \,,
    \end{aligned}
\end{equation}
where $g_{(10)}$ denotes the determinant of the 10D metric. 
The Einstein equation in the presence of a $D0$-brane described by \eqref{SD0} is 
\begin{equation}
\label{EinstD0}
R_{MN} = \frac12 \left(T^{\text{D0}}_{MN}-\frac18 g_{MN} T^{\text{D0}} \right) + \dots \,,  
\end{equation}
where the dots represent terms that are not relevant for the present analysis and $T^{\text{D}0} \equiv g^{MN} T^{\text{D0}}_{MN}$. We choose the static gauge, in which $X^0=\tau$, and assume that the $D0$-brane lies at $x^1=x^2=\dots\,=x^9=0$. Then, from \eqref{TD0MN} it follows that $T^{\text{D0}}_{\mu i} = T^{\text{D0}}_{ij} =0$, so
\begin{equation}
\label{TD0}
T^{\text{D0}} = g^{\mu \nu} T^{\text{D0}}_{\mu \nu} = - \frac{\sqrt{-g_{00}}}{\sqrt{-g_{(10)}}} T_0 e^{-3 \varphi/4} \delta(x^1) \delta^{(8)} (x^i) \, .
\end{equation}
As a result, when we contract the $\mu\nu$-component of equation \eqref{EinstD0} with $g^{\mu \nu}$, we find 
\begin{equation}
\label{RextD0}
R_{2D} = \frac38 T^{\text{D}0}  + \dots \,. 
\end{equation}
This is a 10D equation telling us how the $D0$-brane backreacts on the geometry due to its tension $T_0$. To read off the tension as seen from the point of view of the 2D theory, we have to smear the $D0$-brane along the compact 8D internal space and let it be localized only along the one spatial dimension that is left. This means that the 8D delta function appearing in \eqref{TD0} is replaced by 
\begin{equation}
\label{smear}
 \delta^{(8)} (x^i) \to  \frac{\sqrt{g_{(8)}}}{\text{Vol}[X_8]}  \,, 
\end{equation}
where ${g_{(8)}}=\det g_{ij}$ and $\int_{X_8} d^8 x  \sqrt{g_{(8)}} =  \text{Vol}[X_8]$, such that $\int_{X_8} d^8 x \delta^{(8)}(x^i) = 1$ is preserved. After the replacement \eqref{smear}, equation \eqref{RextD0} becomes
\begin{equation}
    \label{Rextsm}
    R_{2D} = - \frac{3}{8} \frac{(2 \pi)^{7/8}}{\text{vol}[X_8]} \frac{\sqrt{-g_{00}}}{\sqrt{-g_{(2)}}} e^{-3 \varphi/4} \delta(x^1) + \dots \,,
\end{equation}
where $g_{(2)} = \det g_{\mu \nu}$. Thus, we can read off the tension of the smeared $D0$-brane to be
\begin{equation}
\label{TD0sm}
T_\text{smeared D0}  \sim \frac{1}{e^{3 \varphi/4} \text{Vol}[X_8]} \,.
\end{equation}
From the relations \eqref{LAdS2} and \eqref{TD0sm} it then follows that 
\begin{equation}
\label{noscsepD0}
\frac{1}{L} \sim |f_8| T_\text{smeared D0} \gtrsim T_\text{smeared D0} \,,
\end{equation}
which is in accordance with our general argument against scale separation.
This provides a new way to see why this AdS$_2$ vacuum taken from \cite{Lust:2020npd} is not scale separated. In particular, our approach is complementary to that of \cite{Lust:2020npd} since for us scale separation is obstructed due to the presence of a new scale, here $T_{\text{smeared D0}}$, which is not necessarily geometric. Nevertheless, for this specific example it is known that the tension of $D0$-branes has a geometric interpretation in terms of the Kaluza-Klein scale of the eleventh spacetime direction in circle compactifications of 11D supergravity. Hence, in this case, one could say that there is no scale separation also in the traditional geometric sense.

While the $D0$-brane is the most natural candidate, from a bottom-up perspective it is not the only possibility because the BPS domain wall can be engineered in different ways. Indeed, one can notice that by virtue of \eqref{DIL} the external 2D curvature can also be written in terms of the magnetic $F_4$ flux as
\begin{equation} 
\label{2D-EXT}
R_{2D} = - \frac23 e^{\varphi/2} |F_4|^2 \, .
\end{equation}
In the construction of \cite{Lust:2020npd} there are three types of $F_4$ flux depending on which 4-cycle is threaded by the magnetic $F_4$; each of the three 4-cycles is topologically $\rm{S}^2 \times \rm{T}^2$. In \cite{Lust:2020npd} all three possible $F_4$ flux choices are switched on. 
Let us focus here on one of the three 4-cycles which we denote $\Sigma_4$. 
The flux quantization condition for the given 4-cycle is 
\begin{equation}
\int_{\Sigma_4} F_4 = (2 \pi l_s)^3 f_4 \, ,
\end{equation}
where $f_4 \in \mathbb{Z}$. 
Then, taking into account that we are focusing on one 4-cycle, 
we can deduce for the 2D external curvature that  
\begin{equation}
\label{RextS4}
R_{2D} < - \frac23 \left( e^{\varphi/4}  \frac{(2 \pi l_s)^3 f_4}{\text{Vol}[\Sigma_4]} \right)^2 \, .
\end{equation}
The inequality follows from the fact that the contribution from the other $F_4$ fluxes threading the rest of the 4-cycles will only lower further the right hand side. From this inequality we can read off the bound on the length scale of $\rm{AdS}_2$, 
\begin{equation}
    \label{LAdSS4}
    \frac{1}{L} > \frac{1}{\sqrt{3}} \frac{(2 \pi )^{3/8} |f_4|}{\text{Vol}[\Sigma_4]} e^{\varphi/4},
\end{equation}
where we have set $l_s = (2 \pi)^{-7/8}$. 
The electric dual of $F_4$ is $F_6$, which is sourced by a $D4$-brane.  
This $D4$-brane is wrapped on the same 4-cycle as the electric $F_6$ flux, that is a 4-cycle normal to $\Sigma_4$. 
In this way, $F_6$ acts effectively as an $F_2^{eff.}$ flux from a 2D perspective, and the $D4$-brane is effectively a $0$-brane that interpolates between the different values of $F^{eff.}_2$. 
We can then verify that the AdS$_2$ scale is of the same order as the tension of the smeared $D4$-brane by following the same procedure as for the $D0$-brane. 
We will focus of course on the specific 4-cycle $\Sigma_4$ from now on.

We split the coordinates of the 8D internal space as $x^i = (x^{\hat{i}}, x^{\tilde{i}})$, where $x^{\hat{i}}$, $\hat{i}=2,\dots,5$, are the coordinates tangential to $\Sigma_4$ and $x^{\tilde{i}}$, $\tilde{i}=6,\dots,9$, are the internal coordinates that are normal to $\Sigma_4$. 
Then, the spacetime coordinates that are parallel to the $D4$-brane are $(x^0,x^{\tilde{i}})$, while the coordinates $(x^1,x^{\hat{i}})$ are transverse to it. We also assume that the metric of the internal manifold is 
\begin{equation}
    \label{gint}
    ds_8^2 = g_{ij} dx^i dx^j = g_{\hat{i} \hat{j}} d x^{\hat{i}} d x^{\hat{j}} + g_{\tilde{i} \tilde{j}} dx^{\tilde{i}} dx^{\tilde{j}}.
\end{equation} 
The $D4$-brane is described by an action of the form \eqref{SDp} with $p=4$, its worldvolume, $\mathcal{W}_5$, is parametrized by the coordinates $\xi^{\alpha} = (\xi^0, \xi^{\tilde{i}})$ and is described by equations of the form $x^M=X^M(\xi^{\alpha})$. Let the $D4$-brane sit at $x^1=x^{\hat{i}}=0$. Then, from the general formula \eqref{TDpMN} it follows that the non-vanishing components of the energy-momentum tensor associated with the $D4$-brane are 
\begin{align}
    \label{TD4munu}
    T^{\text{D4}}_{\mu \nu} &= \frac{\sqrt{\tilde{g}_{(4)}}}{\sqrt{-g_{00}} \sqrt{-g_{(10)}}} T_4 e^{\varphi/4} g_{0 \mu} g_{0 \nu} \delta(x^1) \delta^{(4)} (x^{\hat{i}}) \,,\\
    \label{TD4ij}
    T^{\text{D4}}_{\tilde{i} \tilde{j}} & = - \frac{\sqrt{-g_{00}}  \sqrt{\tilde{g}_{(4)}}} {\sqrt{-g_{(10)}}} T_4 e^{\varphi/4} g_{\tilde{i} \tilde{j}} \delta(x^1) \delta^{(4)} (x^{\hat{i}}) \,,
\end{align}
where $T_4 = \frac{1}{(2\pi)^4 l_s^5} = (2 \pi)^{3/8}$ and $\tilde{g}_{(4)} \equiv \det g_{\tilde{i} \tilde{j}}$, while its trace is equal to 
\begin{equation}
    \label{TD4}
    T^{\text{D4}} \equiv g^{MN} T^{\text{D4}}_{MN} = - 5 \frac{\sqrt{-g_{00}} \sqrt{\tilde{g}_{(4)}}}{\sqrt{-g_{(10)}}} T_4 e^{\varphi/4} \delta(x^1) \delta^{(4)} (x^{\hat{i}})\, .
\end{equation}
By contracting with $g^{\mu \nu}$ the $\mu \nu$-component of the 10D Einstein equation in the presence of the $D4$-brane, namely
\begin{equation}
\label{EinstD4}
R_{MN} = \frac12 \left(T^{\text{D4}}_{MN}-\frac18 g_{MN} T^{\text{D4}} \right) + \dots \,,  
\end{equation}
we find the backreaction of the $D4$-brane on the external curvature, 
\begin{equation}
\label{RextD4}
R_{2D} = - \frac{1}{40} T^{\rm{D4}} + \dots \,,
\end{equation}
where the dots stand for terms not relevant for the current analysis. 
Then, we smear the $D4$-brane along the internal directions that are transverse to it by the replacement
\begin{equation}
\label{smearD4}
\delta(x^1) \delta^{(4)} (x^{\hat{i}}) \to \delta(x^1)\frac{\sqrt{\hat{g}_{(4)}}}{\text{Vol}[\Sigma_4]} \,,
\end{equation}
where $\hat{g}_{(4)} \equiv \det g_{\hat{i} \hat{j}}$ and we keep the delta function only in the spatial dimension of the external AdS$_2$ spacetime. 
After this replacement, equation \eqref{RextD4} becomes 
\begin{equation}
    \label{RextD4sm}
    R_{2D} = \frac{1}{8} \frac{(2 \pi)^{3/8}}{\text{Vol}[\Sigma_4]} \frac{\sqrt{-g_{00}}}{\sqrt{-g_{(2)}}} e^{\varphi/4} \delta (x^1) + \dots \,,
\end{equation}
so the $D4$-brane backreacts as an effective zero-brane with tension 
\begin{equation}
\label{TD4sm}
T_\text{eff. 0-brane} \sim \frac{ e^{\varphi/4}}{\text{Vol}[\Sigma_4]} \,. 
\end{equation}
Hence, we see that $|R_{2D}| \gtrsim (T_\text{eff. 0-brane})^2$, 
which means that the AdS$_2$ scale is larger or at least of the same order as the tension of the smeared $D4$-brane which, from a 2D perspective, is a domain wall particle.

We observe that the sign of the zero-brane backreaction in \eqref{RextD4sm} is the opposite compared to \eqref{Rextsm}. 
Both cases are consistent as can be verified by the explicit supergravity calculation that leads to \eqref{dilELans}. 
In \eqref{dilELans} we see that $\phi(0)$ being positive or negative does not alter the fact that we can derive a stable BPS solution. 
In addition, for a given sign of $Q$ the sign choice of $\phi(0)$ simply changes the chirality of the preserved Killing spinor as seen by the equations \eqref{eta=-g1eta} and \eqref{g1eta=eta}. 
This means that one sign leads to (1,0) whereas the other to (0,1) from the original (1,1) of the supergravity background. 
In the end this means the sign is just a matter of orientation of the zero-brane.

Before closing this subsection, let us show that the weak coupling and large volume limit implies large flux numbers. 
This implies that the tension of the zero-branes is not only comparable to the background AdS energy scale, but it is actually parametrically smaller. 
Therefore, the AdS background not only probes but also has access to a considerable amount of states within the tower of BPS states. 
Let us see how the large flux appears for the setup of \cite{Lust:2020npd} within our conventions. 
From the expression \eqref{2D-EXT} for the 2D curvature, we have 
\begin{equation}
\frac{1}{L^2} = \frac13 e^{\varphi/2} |F_4|^2  
= \frac{(2 \pi)^{3/4}}{3} e^{\varphi/2} \sum_{\alpha=1}^3 \frac{(f_4^\alpha)^2}{\text{Vol}[\Sigma_4^\alpha]^2} \,, 
\end{equation}
where we have included the sum over all three 4-cycles, while $f_4^\alpha$ is the quantized flux associated with the component of $F_4$ integrated over $\Sigma_4^\alpha$. 
We multiply this equation by the square of one of the volumes of the 4-cycles, say $\Sigma_4^{\hat \alpha}$, that is $(\text{Vol}[\Sigma_4^{\hat \alpha}])^2$, such that we get 
\begin{equation}
\label{VoverL}
\frac{(\text{Vol}[\Sigma_4^{\hat \alpha}])^2}{L^2} 
= \frac{(2 \pi)^{3/4}}{3} e^{\varphi/2} 
\left( (f_4^{\hat \alpha})^2 
+ \sum_{\alpha \ne{\hat \alpha}} (f_4^\alpha)^2 \frac{(\text{Vol}[\Sigma_4^{\hat \alpha}])^2}{(\text{Vol}[\Sigma_4^\alpha])^2} 
\right) \,. 
\end{equation}
Since the 4-cycle $\Sigma_4^{\hat \alpha}$ contains the $S^2$, which has volume $L^2$ (this can be seen from the Einstein equations), the large-volume limit enforces the condition 
\begin{equation}
\frac{(\text{Vol}[\Sigma_4^{\hat \alpha}])^2}{L^2}  \gg 1 \,, 
\end{equation}
taking into account that we are working with the units $l_s = (2 \pi)^{-7/8}$. 
Then, imposing this on \eqref{VoverL}, we deduce that there must be at least one of the three $f_4^\alpha$ that takes parametrically large values. 
If the six-torus is isotropic, then all the $f_4^\alpha$ will be parametrically large. In the anisotropic case, we can still pick that specific $f_4^\alpha$ to be the $f_4$ such that 
\begin{equation}
f_4 \gg 1 \,, 
\end{equation} 
which enters equation \eqref{RextS4}. The above condition tells us that \eqref{eq:defscalesep} is violated parametrically in this example, indeed from \eqref{LAdSS4} we have that $1/L > |f_4| T_\text{eff. 0-brane}  \gg T_\text{eff. 0-brane}$.

\subsection{General argument for flux compactifications} 

The previous discussion involving a specific type IIA example can be generalized to other type II setups in which the AdS$_2$ is supported by some flux. 
Indeed, in this case one typically has
\begin{equation}
R_{2D} \sim - e^{\frac{5-n}{2} \varphi} |F_n|^2 \,, 
\end{equation}
for a magnetic flux $F_n$. 
The precise coefficient of $e^{\frac{5-n}{2} \varphi} |F_n|^2$ is not known \emph{a priori}, because it depends on how the equations of motion relate the various fluxes and also on the possible contributions from the internal space and the sources. 
However, on a flux-supported vacuum we expect the above relation to hold up to some constant coefficient.  
Besides, due to flux quantization we also have 
\begin{equation}
\int_{\Sigma_n} F_n \sim f_n \,, 
\end{equation}
where $f_n \in \mathbb{Z}$. 
Here, $\Sigma_n$ is the $n$-cycle threaded by the magnetic flux $F_n$. 
Once we insert the value of the quantized flux into the external curvature we have 
\begin{equation}
R_{2D} \sim- \left( e^{\frac{5-n}{4} \varphi} \frac{f_n}{\text{Vol}[\Sigma_n]} \right)^2 
\qquad \Rightarrow \qquad 
\frac{1}{L} \sim e^{\frac{5-n}{4} \varphi} \frac{|f_n|}{\text{Vol}[\Sigma_n]} \,. 
\end{equation}

Then, we want to relate this scale to the tension of an effective $0$-brane, which can arise as an appropriately wrapped $Dp$-brane.
The idea is that we turn the magnetic $F_n$ into an electric $F_{10-n}$. 
The latter is sourced by a $Dp$-brane with $p=8-n$ which is electrically coupled to the $C_{9-n}$ potential. 
We split once more the coordinates of the 8D internal space as $x^i = (x^{\hat{i}}, x^{\tilde{i}})$, where $x^{\hat{i}}$, $\hat{i}=2,\dots,n+1$, are the coordinates tangential to $\Sigma_n$ and $x^{\tilde{i}}$, $\tilde{i}=n+2,\dots,9$ are the internal coordinates that are normal to $\Sigma_n$. 
The spacetime coordinates that are parallel to the $Dp$-brane are $(x^0,x^{\tilde{i}})$, while the coordinates $(x^1,x^{\hat{i}})$ are transverse to it. We also assume that the metric of the internal manifold is split as in \eqref{gint}. 
Taking without loss of generality the $D(8-n)$-brane source to lie at $x^1=x^{\hat{i}}=0$, its backreaction on the 2D external curvature is
\begin{equation}
\label{RextD}
R_{2D} = \frac{1}{8} (5-n) (2 \pi)^{\frac{n-1}{8}} e^{\frac{5-n}{4} \varphi} 
\frac{\sqrt{-g_{00}} \sqrt{\tilde{g}_{(8-n)}}}{\sqrt{-g_{(10)}}} \delta(x^1) \delta^{(n)} (x^{\hat{i}}) + \dots \,. 
\end{equation}
The coefficient $\frac{1}{8} (5-n) (2 \pi)^{\frac{n-1}{8}}$ on the right-hand side of the last equation vanishes for $n=5$, that is for a D3-brane in type IIB, while it is of the order of 1 for all other $0\leq n \leq 8$. 
This means that in a IIB setup one should use the other Dp-branes (e.g. D7) and not the D3 to realize our argument in search for towers of light superparticles. 

Next, we smear the $D(8-n)$-brane along the internal $n$-cycle (such that it looks like a particle in 2D) by replacing
\begin{equation}
\delta(x^1) \delta^{(n)} (x^{\hat{i}}) \to \delta(x^1)\frac{\sqrt{\hat{g}_{(n)}}}{\text{Vol}[\Sigma_n]}  \,. 
\end{equation}
Inserting this into \eqref{RextD} we can deduce the effective mass of the particle from its gravitational backreaction, which reads 
\begin{equation}
T_\text{eff. 0-brane} \sim  \frac{ e^{\frac{5-n}{4} \varphi}}{\text{Vol}[\Sigma_n]} \,. 
\end{equation}
Hence, these flux-supported AdS$_2$ vacua in type II compactifications are not scale separated, for they violate \eqref{eq:defscalesep}. Indeed, we generically expect light 0-branes to be present in the bulk, with their tension being parametrically of the same order as the AdS$_2$ scale. Notice again that, while $T_\text{eff. 0-brane} > \Lambda_{UV}$, the scale $\Lambda_{UV}$ does not need to be geometric for the purposes of our argument. Thus, our results exclude scale separation in a broader sense than what usually achieved. 

Note that even though we worked here with type II, clearly the same analysis carries through also for heterotic and type I theories, 
as long as one compactifies on a background with enough supersymmetry to support BPS objects.

\section{Discussion}

In this work, we provided a bottom-up argument excluding the existence of AdS$_2$ vacua supported by fluxes and with at least two preserved supercharges. This argument is specific to 2D and does not have an obvious (to us) generalization to higher dimensions. It relies on the existence of fundamental BPS domain walls interpolating between two such vacua and whose tension gives an upper bound on the UV cutoff of the 2D effective theory. 

Previous investigations \cite{Cribiori:2022trc,Cribiori:2023ihv} could rule out scale separation only up to four supercharges. In this sense, the present work represents an improvement on the state of the art. Furthermore, since we deal with supersymmetric vacua, we believe that, when protected by enough preserved supercharges, our results can be extrapolated beyond perturbation theory and thus be relevant away from the supergravity approximation.

A direct consequence of our results is that, if scale separation is possible at all in 2D, it most likely occurs for $\mathcal{N}\leq(1,0)$ AdS$_2$ vacua. These are not so straightforward to construct, but one possibility is to compactify type II superstring theory on a manifold with $Spin(7)$-holomony, which indeed preserves 1/16 of the original supersymmetry, and then perform an appropriate orientifold projection to reduce it further by a factor 1/2. The non-trivial bit is to satisfy the tadpole cancellation condition while preserving the remaining amount of supersymmetry; besides, one still has to further check that indeed the background will not have light branes. 

Another open question concerns the fate of dS$_2$ vacua, which in higher dimensions can usually be constrained with similar tools as for scale separation. It would be interesting to understand if this is the case also in 2D or if, on the contrary, dS$_2$ vacua are somehow peculiar. While we leave a proper analysis for future work, what we can immediately state here is that a model with pure 2D supergravity coupled to a goldstino will be ruled out by our argument. 
Indeed, to break supersymmetry and uplift to de Sitter will require an effective contribution to the Ricci scalar of the form $R \sim M^2+\dots $, where the dots contain derivative and/or fermionic couplings due to the 2D Volkov--Akulov system. 
Since $M \sim \Lambda_{UV}$ is the non-linear supersymmetry breaking scale and since the cosmological constant is $1/L_H^2 \sim H^2 \sim M^2$, we will find that $L_{H} \Lambda_{UV} \sim 1$ in this case as well. 
Once more, this argument is particular to 2D. 
Note that for 4D and 5D supergravity with at least eight preserved supercharges at the Lagrangian level, arguments based on the weak gravity conjecture have been proposed to restrict dS vacua, due to the difficulty of achieving scale separation \cite{Cribiori:2020wch,Cribiori:2020use,DallAgata:2021nnr,Cribiori:2023ihv}.

One can also investigate what happens for the case of the supersymmetric de Sitter algebra which can deliver consistent theories only in 2D, see for example \cite{Anninos:2023exn}. 
We expect that an argument similar to our supersymmetric AdS$_2$ analysis also holds there, that is unless supersymmetry is minimal, the system cannot be scale separated. 
If supersymmetry is minimal, we cannot at this point exclude scale separation.

\section*{Acknowledgments}

We would like to thank George Tringas, Vincent Van Hemelryck and Thomas Van Riet for discussions. 
The work of NC is supported by the Research Foundation Flanders (FWO grant 1259125N). 
The research work was supported by the Hellenic Foundation for Research and Innovation (HFRI) under the 3rd Call for HFRI PhD Fellowships (Fellowship Number: 6554). 
\begin{center}
\includegraphics[scale=0.35]{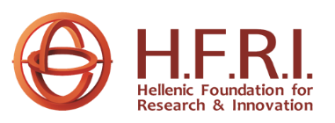}
\end{center}

\appendix

\section{2D $\mathcal{N}=(1,1)$ supergravity, superparticle and one-form multiplets}
\label{eq:SUGRAconv}

In this appendix, we collected the basic tools and ingredients needed to construct the 2D $\mathcal{N}=(1,1)$ supergravity models employed in the main part of the article.

\subsection{Gamma matrices and spinors in 2D}
\label{AppA}

For gamma matrices and spinors in 2D, we follow the notation and conventions of Chapter 3 of \cite{Freedman:2012zz}. The gamma matrices, $\gamma_a$, satisfy the Clifford algebra
\begin{equation}
    \label{Cliff2D}
    \{ \gamma_a, \gamma_b \} = 2 \eta_{ab} {\mathbb 1}_{2} ,
\end{equation}
where $\eta_{ab} = \text{diag}(-1,1)$. The indices $a,b=0,1$ are flat, while $\mu,\nu=0,1$ are for curved spacetime. We also define 
\begin{align}
\label{gamma3}
& \gamma_3 \equiv - \gamma_0 \gamma_1  , \\
\label{gammaab}
& \gamma_{ab} \equiv \gamma_{[a} \gamma_{b]} = - \epsilon_{ab} \gamma_3 \,, 
\end{align}
where $\epsilon_{ab}$ is the 2D antisymmetric Levi-Civita symbol with $\epsilon_{01} = 1$.
The charge conjugation matrix, $C$, satisfies the relations 
\begin{align}
    \label{Cunit}
     C^{\dagger} C &= {\mathbb 1}_{2}  \,, \\
    \label{Cantisymm}
     C^T & = - C  \,, \\
    \label{gammaT}
    \gamma_a^T &= - C \gamma_a C^{-1}, \\
    \label{gammaabT}
    \gamma_{ab}^T &= - C \gamma_{ab} C^{-1} . 
\end{align}
Given any two-component spinor $\lambda$ in two dimensions, we define its Majorana conjugate as 
\begin{equation}
    \label{Majconj}
    \bar{\lambda} = \lambda^T C \, .
\end{equation}
From the definition \eqref{Majconj} and the properties \eqref{Cantisymm}-\eqref{gammaabT}, it follows then that 
\begin{equation}    \label{Majflips}
	\bar{\lambda} M \chi =   \left\{ 
	\begin{array}{l}
		+ \bar{\chi} M \lambda \qquad \textrm{for } M={\mathbb 1}_{2}, \\[4mm]
		- \bar{\chi} M \lambda\qquad \textrm{for } M=\gamma_{a}, \gamma_{ab} \,,
	\end{array}
	\right.
\end{equation}
for any two spinors $\lambda$ and $\chi$ with anticommuting components. 
Furthermore, the following Fierz identity holds,
\begin{equation}
    \label{Fierz}
    \chi \bar{\lambda} = - \frac{1}{2} (\bar{\lambda} \chi) {\mathbb 1}_2 - \frac{1}{2} (\bar{\lambda} \gamma^a \chi) \gamma_a - \frac{1}{2} (\bar{\lambda} \gamma_3 \chi) \gamma_3 \, .
\end{equation}
The charge conjugate of any spinor $\lambda$ is defined by 
\begin{equation}
    \label{chconj}
    \lambda^{c} = B^{-1} \lambda^*, 
\end{equation}
where $B=i C \gamma^0$. A spinor is Majorana if it is equal to its own charge conjugate:
\begin{equation}
\text{Majorana}: \quad    \psi = \psi^c \quad \Longleftrightarrow  \quad \psi^* = B \psi \, . 
\end{equation}
After introducing the chirality projectors
\begin{equation}
    P_L \equiv \frac{1}{2} \left( \mathbb{1}_2 + \gamma_3 \right) , \qquad  P_R \equiv \frac{1}{2} \left( \mathbb{1}_2 - \gamma_3 \right),
\end{equation}
we can define a left-handed Weyl spinor $\chi$ and right-handed Weyl spinor $\xi$  as the spinors satisfying the conditions
\begin{align}
P_L \chi = \chi \quad \iff \quad \gamma_3 \chi &= \chi \, ,  \\
    P_R \xi = \xi \quad \iff \quad \gamma_3 \xi &= - \xi \, . 
\end{align}
Since $(\gamma_3 \psi)^c = \gamma_3 \psi^c$, the left- and right-handed projections of a Majorana spinor $\psi$, namely $\psi_L \equiv P_L \psi$ and $\psi_R \equiv P_R \psi$ respectively, are also Majorana and are thus called Majorana-Weyl spinors. They are irreducible spinors in 2D.

\subsection{2D $\mathcal{N}=(1,1)$ supergravity multiplets and action}

The central ingredient of 2D $\mathcal{N}=(1,1)$ supergravity is the gravity multiplet, which consists of a zweibein $e_{\mu}^a\,,$ a Majorana gravitino $\psi_{\mu}$ and an auxiliary real scalar $A$. 
The local supersymmetry transformation rules for these fields can be found in \cite{Howe:1978ia}; in our conventions, they read  
\begin{align}
\label{dema}
&\delta_{\epsilon} e_{\mu}^a  = \frac{1}{2} \bar{\epsilon} \gamma^a \psi_{\mu}\, ,\\
\label{dpsimu}
& \delta_{\epsilon} \psi_{\mu}  = D_{\mu} \epsilon + \frac{1}{2} A \gamma_{\mu} \epsilon \, , \\
\label{dA}
&\delta_{\epsilon} A  = - \frac{1}{2} \epsilon^{\mu \nu} \bar{\epsilon} \gamma_3 {\hat D}_{\mu} \psi_{\nu} \, ,
\end{align}
where $\epsilon$ is a spacetime-dependent Majorana spinor, $D_{\mu}$ denotes the covariant derivative with respect to local Lorentz transformations, for example, 
\begin{equation}
\label{Dmue}
D_{\mu} \epsilon = \partial_{\mu} \epsilon + \frac{1}{4} \omega_{\mu a b} \gamma^{ab} \epsilon = \partial_{\mu} \epsilon + \frac{1}{2} \omega_{\mu} \gamma_3 \epsilon \, , 
\end{equation}
and 
\begin{equation}
\label{Dhatpsi}
{\hat D}_{\mu} \psi_{\nu} \equiv D_{\mu} \psi_{\nu} - \frac{1}{2} A \gamma_{\nu} \psi_{\mu} \,. 
\end{equation}
The spin connection is $\omega_{\mu ab} = \epsilon_{ab} \omega_{\mu}$, where $\omega_{\mu} \equiv - \frac{1}{2} \epsilon^{ab} \omega_{\mu ab} = e_{\mu}^a \epsilon^{\nu   \rho} \partial_{\nu} e_{a \rho} - \frac{1}{4} \epsilon^{\nu \rho} \bar{\psi}_{\nu} \gamma_{\mu} \psi_{\rho} $.
A second important ingredient is the 2D $\mathcal{N}=(1,1)$ scalar multiplet, which comprises a real scalar field $\phi$, a Majorana fermion $\lambda$ and an auxiliary real scalar $F$. The local supersymmetry transformation rules for such a mulitplet coupled to supergravity are \cite{West:1986wua}
\begin{align}
\label{dephi}
& \delta_{\epsilon} \phi = \frac{1}{2} \bar{\epsilon} \lambda \, , \\
\label{delambda}
& \delta_{\epsilon} \lambda = \frac{1}{2} \gamma^{\mu} \epsilon {\hat D}_{\mu} \phi + \frac{1}{2} F \epsilon \, , \\
\label{deF}
& \delta_{\epsilon} F = \frac{1}{2} \bar{\epsilon} \gamma^{\mu} {\hat D}_{\mu} \lambda \, ,
\end{align}
where 
\begin{align}
\label{hatDphi}
& {\hat D}_{\mu} \phi \equiv \partial_{\mu} \phi - \frac{1}{2} \bar{\psi}_{\mu} \lambda \, ,\\
\label{hatDlam}
& {\hat D}_{\mu} \lambda \equiv D_{\mu} \lambda - \frac{1}{2} \gamma^{\nu} \psi_{\mu} {\hat D}_{\nu} \phi - \frac{1}{2} F  \psi_{\mu} \,.
\end{align}
The local supersymmetry transformations of these multiplets obey the algebra
\begin{equation}
\label{QQcom}
\left[\delta_Q (\epsilon_1),\delta_Q (\epsilon_2)\right] = \delta_{\text{gct}} (\xi^{\mu}) - \delta_{\text{Lorentz}} (\xi^{\mu} \omega_{\mu}) - \delta_Q (\xi^{\mu} \psi_{\mu}) + \delta_{\text{Lorentz}} \left( - \frac{1}{2} A \bar{\epsilon}_1 \gamma_3 \epsilon_2 \right)  ,
\end{equation}
where the first term on the right-hand side denotes a general coordinate transformation with diffeomorphism parameter $\xi^{\mu} = - \frac{1}{2} \bar{\epsilon}_1 \gamma^{\mu} \epsilon_2 $.

The $\mathcal{N}=(1,1)$ supersymmetric extension of Jackiw-Teitelboim (JT) gravity \cite{Teitelboim:1983ux,Jackiw:1984je} contains the 2D $\mathcal{N}=(1,1)$ supergravity multiplet coupled to a scalar multiplet $(\phi,\lambda,F)$, called the dilaton multiplet. The locally supersymmetric action describing the system is given, up to a dimensionless real proportionality constant, by \cite{Chamseddine:1991fg,Astorino:2002bj}
\begin{equation}
\begin{aligned}
    \label{JTSG}
    S = \int d^2 x \, e &  \left( \phi R  - 2 \left( A + \frac{1}{L} \right) F - \frac{2}{L} \phi A + \frac{1}{2L} \phi \epsilon^{\mu \nu}  \bar{\psi}_{\mu} \gamma_3 \psi_{\nu}  \right. \\ & \quad \left.- 2 \epsilon^{\mu \nu} \bar{\lambda} \gamma_3 D_{\mu} \psi_{\nu} + \frac{1}{L} \bar{\lambda} \gamma^{\mu} \psi_{\mu} \right) , 
\end{aligned}
\end{equation}
where $e=\text{det}(e_{\mu}^a)$, $R=2 \epsilon^{\mu \nu} \partial_{\mu} \omega_{\nu}$ is the 2D Ricci scalar, and $L$ is a real constant with dimension of length. 
Varying the action \eqref{JTSG} with respect to the component fields of the dilaton multiplet we obtain
\begin{align}
    \label{Feom-d}
    \delta F & : \ \ A = - \frac{1}{L} \,, \\ 
    \label{lameom}
    \delta \lambda & : \ \ \epsilon^{\mu \nu} \hat{D}_{\mu} \psi_{\nu} = 0 \, , \\
    \label{phieom}
    \delta \phi & : \ \ R = - \frac{2}{L^2} - \frac{1}{2L} \epsilon^{\mu \nu} \bar{\psi}_{\mu} \gamma_3 \psi_{\nu} \, , 
    \end{align}
where for the derivation of \eqref{lameom} and \eqref{phieom} we have used \eqref{Feom-d} and the definition \eqref{Dhatpsi}. In particular, equation \eqref{phieom} implies an $\text{AdS}_2$ vacuum, with $L$ identified with the corresponding AdS radius, given that $R_{\mu \nu \rho \sigma} = R \, g_{\rho [\mu} g_{\nu] \sigma}$ in 2D.

\subsection{Superparticle coupled to the one-form dilaton multiplet}
\label{particledil}

In this appendix, we provide the action that describes the coupling of a superparticle in a curved 2D $\mathcal{N}=(1,1)$ superspace to the one-form dilaton multiplet. This superparticle is the BPS domain wall in the main text.

Let ${\cal M}^{2|2}$ denote the 2D $\mathcal{N}=(1,1)$ superspace, which is parametrized by the supercoordinates $z^M = (x^{\mu},\theta^{\alpha})$, where $\theta^{
\alpha
}$, $\alpha=1,2$, are the anticommuting components of a Majorana spinor $\theta$. We consider a superparticle with worldline ${\cal W}_1$ embedded into ${\cal M}^{2|2}$ and parametrized by the coordinate $\tau$. The worldline ${\cal W}_1$ is described by parametric equations of the form
\begin{equation}
    \label{zM(tau)}
    z^M = Z^M (\tau) = (X^{\mu} (\tau), \Theta^{\alpha} (\tau)) \,.
\end{equation}
We then introduce the supervielbein one-forms $E^A = E^A_M (z) dz^M =  (E^a, \Psi^{\alpha})$, where $E^a$ is a bosonic super-one-form and $\Psi$ is a Majorana spinor-valued one. The projections of the supervielbein on the spacetime submanifold of ${\cal M}^{2|2}$, that is the $\theta = d \theta = 0 $ hypersurface, are given by  
\begin{equation}
    \label{supervielbeinproj}
    E^a |_{\theta = d \theta = 0} = e^a_{\mu} (x) dx^{\mu}, \, \Psi|_{\theta = d \theta = 0} = \psi_{\mu} (x) dx^{\mu} .  
\end{equation}
Furthermore, the pullback of the bosonic supervielbein $E^a$ on ${\cal W}_1$ equals
\begin{equation}
    \label{pullEa}
    {\hat E}^a \equiv E^a_M(Z(\tau))  d Z^M (\tau) = E_{\tau}^a (\tau) d \tau \,,
\end{equation}
where 
\begin{equation}
    \label{Eta}
    E_{\tau}^a (\tau) \equiv E^a_M(Z(\tau)) \dot{Z}^M (\tau) \, ,
\end{equation}
where the dot denotes a derivative with respect to $\tau$. 

Let $\Phi$ denote the one-form dilaton superfield, whose lowest component is $\Phi(z)|_{\theta = 0} = \phi(x)$, while its exterior derivative reads
\begin{equation}
\label{dPhi}
d \Phi = (d \Phi)_a E^a + \frac{1}{2} \bar{\Psi} \Lambda \,,
\end{equation}
where $\Lambda$ is a Majorana spinor superfield with lowest component $\Lambda|_{\theta=0}=\lambda$. 
The action describing the coupling of the superparticle to the one-form dilaton multiplet is 
\begin{equation}
    \label{Ssp}
    S_{\text{sp}} = - |Q| \int_{{\cal W}_1} d \tau \sqrt{-h} |\hat{\Phi}| + Q \int_{{\cal W}_1} \hat{\cal B}_1 \,,
\end{equation}
where
\begin{equation}
    \label{h(t)}
    h(\tau) \equiv \eta_{ab} E^a_{\tau} (\tau) E^b_{\tau} (\tau)
\end{equation}
is the induced metric on ${\cal W}_1$, while $\hat{\Phi} (\tau) \equiv \Phi (Z(\tau))$ and $\hat{\cal B}_1$ are the pullbacks of $\Phi$ and a super-one-form ${\cal B}_1$ respectively on ${\cal W}_1$. The projection of ${\cal B}_1$ on spacetime is the one-form $B_1$ introduced in equation \eqref{starH2main}, while its super-field strength equals 
\begin{align}
    \label{H2sup}
    {\cal H}_2 = d {\cal B}_1 = \frac{1}{2} {\cal H}_{ab} E^a \wedge E^b - \frac{1}{2} \bar{\Lambda} \gamma_a \gamma_3 \Psi \wedge E^a - \frac{1}{4} \Phi \bar{\Psi} \wedge \gamma_3 \Psi . 
\end{align}
Also, $Q$ denotes the real charge of the superparticle under the super-one-from $\mathcal{B}_1$, and sets its tension as $T=|Q|$.

The action \eqref{Ssp} is invariant under a $\kappa$-symmetry transformation parametrized by a Majorana spinor $\kappa(\tau)$ and acting on the pullback of $E^a$ on ${\cal W}_1$ as 
\begin{equation}
    \label{dkappaE}
    \delta_{\kappa} {\hat E}^a = \frac{1}{2} \bar{\kappa} \gamma^a \hat{\Psi} \,, 
\end{equation}
where $\hat{\Psi}$ denotes the pullback of the fermionic supervielbein $\Psi$ on ${\cal W}_1$. Also, the contraction operator along $\kappa(\tau)$, $i_{\kappa}$, acts on the supervielbein as 
\begin{equation}
    \label{ikEA}
    i_{\kappa} \hat{E}^a = 0 \,, \qquad i_{\kappa} \hat{\Psi} = \kappa \, , 
    \end{equation}
and the local fermionic parameter $\kappa(\tau)$ satisifes the projection condition 
\begin{equation}
    \label{projPhi}
    \kappa = \frac{|Q \hat{\Phi}|}{Q \hat{\Phi}} \Gamma \kappa \,,
\end{equation}
where 
\begin{equation}
    \label{Gamma}
    \Gamma \equiv \frac{1}{\sqrt{-h}} E^a_{\tau} \gamma_3 \gamma_a \,,
\end{equation}
which is such that $\Gamma^2 = \mathbb{1}_2$.

\subsection{One-form supergravity multiplet}
\label{app:1formSUGRA}

In this appendix we show that the strategy of dualizing the real auxiliary scalar of the 2D $\mathcal{N}=(1,1)$ supergravity multiplet to a two-form does not lead to the desired effect of producing a dynamical AdS length scale. This is the reason why in the main text we dualized the auxiliary field of the dilaton multiplet instead.

To construct a one-form supergravity multiplet, we introduce a two-form $J_2$, which is locally equal to $J_2 = d A_1$ for some one-form $A_1$ ($d J_2 =0$ in 2D), and write
\begin{equation}
\label{starJ2}
\star J_2 = \frac{1}{2} \epsilon^{\mu \nu} J_{\mu \nu} = \epsilon^{\mu \nu} \partial_{\mu} A_{\nu} = A - \frac{1}{4} \epsilon^{\mu \nu} \bar{\psi}_{\mu} \gamma_3 \psi_{\nu} \,, 
\end{equation}
or, equivalently, 
\begin{equation}
\label{Jmunu}
J_{\mu \nu} = - A \epsilon_{\mu \nu} - \frac{1}{2} \bar{\psi}_{\mu} \gamma_3 \psi_{\nu} \,. 
\end{equation}
Taking the supersymmetry variation of the above equation one obtains an integrable equation that gives the following local supersymmetry transformation rule for the one-form $A_1$:
\begin{equation}
\label{dCm}
\delta_{\epsilon} A_{\mu} = - \frac{1}{2} \bar{\epsilon} \gamma_3 \psi_{\mu} \, .
\end{equation}
Therefore, the commutator of two local supersymmetry transformations acting on $A_{\mu}$ reads 
\begin{equation}
\label{QQcomCm}
[\delta_Q(\epsilon_1),\delta_Q(\epsilon_2)] A_{\mu} = \delta_{\text{gct}}(\xi^{\nu}) A_{\mu} - \delta_{\text{gauge}} (\xi^{\nu} A_{\nu}) A_{\mu}  - \delta_Q (\xi^{\nu} \psi_{\nu}) A_{\mu} + \delta_{\text{gauge}} \left( 
\frac{1}{2} \bar{\epsilon}_1 \gamma_3 \epsilon_2  \right) A_{\mu} \, , 
\end{equation}
where $\delta_{\text{gauge}}$ is an abelian gauge transformation that acts on $A_{\mu}$ as $\delta_{\text{gauge}} (\Lambda) A_{\mu} = \partial_{\mu} \Lambda \,$, where $\Lambda$ is an infinitesimal real parameter. 
Substituting equation \eqref{starJ2} into the action \eqref{JTSG}, the latter becomes 
\begin{equation}
\begin{aligned}
    \label{1formSUGRA}
    S_{\text{1-form SUGRA}} =  \int d^2 x \, e \bigg{[} &\phi R - 2 F \star J_2 - \frac{2}{L} (F + \phi \star J_2) - \frac{1}{2} F \epsilon^{\mu \nu} \bar{\psi}_{\mu} \gamma_3 \psi_{\nu}  \\ 
    & - 2 \epsilon^{\mu \nu} \bar{\lambda} \gamma_3 D_{\mu} \psi_{\nu} + \frac{1}{L} \bar{\lambda} \gamma^{\mu} \psi_{\mu} \bigg{]} . 
    \end{aligned}
    \end{equation}

The action that describes the coupling of a superparticle in a curved 2D $\mathcal{N}=(1,1)$ superspace to the one-form supergravity multiplet is 
\begin{align}
    \label{sp'}
    S_{\text{sp}}' = - |Q| \int_{{\cal W}_1} d \tau \sqrt{-h} + Q \int_{{\cal W}_1} \hat{\cal A}_1\, ,
\end{align}
 where $\mathcal{W}_1$ denotes the worldline of the superparticle, which is embedded into the superspace ${\cal M}^{2|2}$ and described by parametric equations of the form \eqref{zM(tau)}, and $h$ is again the induced metric on $\mathcal{W}_1$, given by \eqref{h(t)}. Also, $\hat{\cal A}_1$ is the pullback of a super-one-form ${\cal A}_1$ on ${\cal W}_1$. The restriction of ${\cal A}_1$ to 2D spacetime is the one-form $A_1$ introduced in \eqref{starJ2} and its field strength reads
\begin{equation}
    \label{superJ2}
    {\cal J}_2 = d {\cal A}_1 = \frac{1}{2} {\cal J}_{ab} E^a \wedge E^b - \frac{1}{4} \bar{\Psi} \wedge \gamma_3 \Psi .
\end{equation}
Moreover, $Q$ is a real constant charge that characterizes the coupling of the superparticle to the super-one-form ${\cal A}_1$ and also determines the particle tension, $T=|Q|$.
The action \eqref{sp'} is invariant under a $\kappa$-symmetry transformation described by equations \eqref{dkappaE} and \eqref{ikEA}, with the Majorana spinor parameter $\kappa(\tau)$ obeying the condition
\begin{equation}
    \label{kproj}
    \kappa = \frac{|Q|}{Q} \Gamma \kappa \, ,
\end{equation}
where the matrix $\Gamma$ is given by \eqref{Gamma}.

Let us now consider the action that is the sum of \eqref{1formSUGRA} and \eqref{sp'}. In a background with vanishing fermionic fields, the Euler-Lagrange equation for the auxiliary scalar of the dilaton multiplet, $F$, is 
\begin{equation}
    \label{Feom}
    \star J_2 = - \frac{1}{L} \,,
\end{equation}
while the corresponding equation for the dilaton $\phi$ reads
\begin{equation}
    \label{R=-2/L2}
    R = \frac{2}{L} \star J_2 \overset{\eqref{Feom}}{=}  - \frac{2}{L^2} \,. 
\end{equation}
It is clear that the value of the flux $J_2$, which determines the cosmological constant, is fixed by the equation of motion for $F$. Therefore, if we replace the auxiliary scalar of the supergravity multiplet with a two-form $J_2 = d A_1$, the length scale of the background $\text{AdS}_2$ spacetime cannot change even in the presence of a particle that is charged under the one-form $A_1$.

\subsection{Derivation of the domain wall solutions}
\label{SolutionofODEs}

In this appendix, we provide more details on the derivation of the domain wall solutions given in subsection \ref{domainwall}. We first solve the ordinary differential equations \eqref{dilELans}-\eqref{Einsxxans} away from the particle, that is for $x\neq0$, where $\delta(x)$ vanishes. For $x\neq0$, equation \eqref{BELxcomp} implies that 
\begin{equation}
\label{B'const}
    \phi^{-1} e^{-f} B' = \left\{ \begin{array}{ll}
         \alpha_{+} &   x>0 \\
        \alpha_{-}  &  x<0   
        \end{array} \right. ,
\end{equation}
where $\alpha_{\pm}$ are real constants. Then, from \eqref{Einsttans} it follows that 
\begin{align}
\label{phi''}
    \phi^{''} - \frac{1}{4} \alpha_{\pm}^2 \phi = 0 \, ,
\end{align}
where the subscripts $+$ and $-$ refer to the regions $x>0$ and $x<0$ respectively. We will focus on the case in which $\alpha_{\pm}$ are both non-zero. A vanishing value for $\alpha_{+}$ or $\alpha_{-}$ implies that $\star H_2$ vanishes for $x>0$ or $x<0$ respectively, as can be seen from equation \eqref{fluxans}, thus resulting in a Minkowski solution in the respective region, as follows from the equation of motion for the dilaton, $\eqref{dilELsource}$. On the other hand, if neither of $\alpha_{\pm}$ is equal to zero, there exists an $\rm{AdS}_2$ solution in each of the regions $x>0$ and $x<0$. In this case, the general solution of \eqref{phi''} is
\begin{equation}
    \label{phisol}
    \phi (x)= \left\{  \begin{array}{ll} \gamma_{+} e^{\frac{1}{2} \alpha_{+} x} + \delta_{+} e^{-\frac{1}{2} \alpha_{+} x} & x \geq 0 \\
     \gamma_{-} e^{\frac{1}{2} \alpha_{-} x} + \delta_{-} e^{-\frac{1}{2} \alpha_{-} x} & x < 0
    \end{array}  \right. , 
\end{equation}
where $\gamma_{\pm}$ and $\delta_{\pm}$ are real constants. Then, given \eqref{B'const} and \eqref{phi''}, the differential equation \eqref{Einsxxans} implies that 
\begin{equation}
    \label{f'phi''}
    \phi^{''} - f' \phi'  = 0   \Longleftrightarrow (e^{-f} \phi')' =0 \,,
    \end{equation}
which is solved by 
\begin{equation}
    f = \ln \left(\frac{\phi'}{c_{\pm}}\right) \, ,
\end{equation}
where $c_{\pm}$ are real constants of integration. Using \eqref{phisol} and setting $\beta_{\pm} = 
\frac{\alpha_{\pm}}{2 c_{\pm}}  $ we find 
\begin{equation}
    \label{fgensol}
    f = \left\{  \begin{array}{ll} \ln [ \beta_{+} (\gamma_{+} e^{\frac{1}{2} \alpha_{+} x} -  \delta_{+} e^{-\frac{1}{2} \alpha_{+} x} )  ]  & x \geq 0 \\ 
    \ln [ \beta_{-} (\gamma_{-} e^{\frac{1}{2} \alpha_{-} x} -  \delta_{-} e^{-\frac{1}{2} \alpha_{-} x} ) ]  & x < 0
    \end{array}  \right. .
\end{equation}
Furthermore, from equations \eqref{B'const}, \eqref{phisol} and \eqref{fgensol} it follows that 
\begin{equation}
    \label{B'sol}
    B' = \alpha_{\pm} \beta_{\pm} (\gamma_{\pm}^2 e^{\alpha_{\pm}x} - \delta_{\pm}^2 e^{-\alpha_{\pm}x}) \,,
\end{equation}
so $B(x)$ is given by 
\begin{equation}
\label{Bsol}
    B = \left\{  \begin{array}{ll} \beta_{+} (\gamma_{+}^2 e^{\alpha_{+}x} +  \delta_{+}^2 e^{-\alpha_{+}x}) & x \geq 0 \\
      \beta_{-} (\gamma_{-}^2 e^{\alpha_{-}x} +  \delta_{-}^2 e^{-\alpha_{-}x}) & x < 0
      \end{array} \right. 
\end{equation}
up to some real constants of integration which can be set equal to zero by a gauge transformation. In addition, the functions \eqref{phisol}, \eqref{fgensol} and \eqref{Bsol} satisfy the differential equation \eqref{dilELans} for $x \neq 0$ as well. 

We also require that the functions $\phi(x)$, $f(x)$ and $B(x)$ be continuous at $x=0$, which leads to the conditions
\begin{align}
    \label{phicont}
    \gamma_{+} + \delta_{+} & = \gamma_{-} + \delta_{-}\,, \\
    \label{fcont}
    \beta_{+} (\gamma_{+} - \delta_{+}) & = \beta_{-} (\gamma_{-} - \delta_{-}) \,, \\
    \label{Bcont}
    \beta_{+} (\gamma_{+}^2 + \delta_{+}^2) & = \beta_{-} (\gamma_{-}^2 + \delta_{-}^2) \,. 
\end{align}
Moreover, $\phi(x)$ and $g_{tt} (x) = - e^{2f (x)}$ must not vanish for any value of $x$. In particular, $\phi(0)$ and $g_{tt}(0)$ are different from zero if $\gamma_{+} + \delta_{+} \neq 0$ and $\gamma_{+}- \delta_{+} \neq 0$ respectively. Then, equations \eqref{phicont}-\eqref{Bcont} allow us to express the parameters $\beta_{-}$, $\gamma_{-}$ and $\delta_{-}$ in terms of $\beta_{+}$, $\gamma_{+}$ and $\delta_{+}$ according to 
\begin{equation}
    \label{minusconst}
    \beta_{-} = \frac{(\gamma_{+} -\delta_{+})^2}{(\gamma_{+}+\delta_{+})^2} \beta_{+} \,, \qquad \gamma_{-} = \frac{\gamma_{+}(\gamma_{+} + \delta_{+})}{\gamma_{+} - \delta_{+}} \,, \qquad \delta_{-} = - \frac{\delta_{+} (\gamma_{+} + \delta_{+})}{\gamma_{+} - \delta_{+}} \,.
\end{equation}

On the other hand, the presence of the delta function $\delta(x)$ on the right-hand side of \eqref{BELxcomp} means that the function $\phi^{-1} e^{-f} B'$ has a discontinuity at $x=0$. In order to compute the difference between its constant values for $x>0$ and $x<0$, $\alpha_{+}$ and $\alpha_{-}$ respectively, we integrate equation \eqref{BELxcomp} over an interval $[-\epsilon, \epsilon]$, where $\epsilon$ is a positive real number. We find
\begin{equation}
    \label{Da}
    \alpha_{+} - \alpha_{-} = - Q \Longleftrightarrow \alpha_{-} = \alpha_{+} + Q \,. 
\end{equation}
Then, we substitute equation \eqref{Einsxxans} into \eqref{Einsttans} and multiply the resulting equation by $e^{-f} \phi$. We obtain
\begin{equation}
    \label{phi''f'delta}
    (e^{-f} \phi')' = - \frac{1}{2} |Q \phi (0)| e^{-f(0)} \delta(x) \,. 
 \end{equation}
 Integrating the last equation over the interval $[-\epsilon, \epsilon]$ and using the formulae \eqref{phisol}, \eqref{fgensol} and \eqref{minusconst} we find 
 \begin{equation}
     \label{gda1}
     4 \gamma_{+} \delta_{+} \alpha_{+} = |\gamma_{+} + \delta_{+}| (\gamma_{+} - \delta_{+})  |Q|-  (\gamma_{+} + \delta_{+})^2 Q\,.
 \end{equation}
 Furthermore, substituting \eqref{Einsxxans} into the differential equation \eqref{dilELans} and multiplying the resulting equation by $\phi^{-1} e^{f}$ gives
 \begin{equation}
     \label{eff'/phi}
     \left( \frac{e^f f'}{\phi} \right)' = - \frac{1}{2} |Q| \frac{e^{f(0)}}{|\phi(0)|} \delta(x) \,.
 \end{equation}
Integrating the above equation over $[-\epsilon, \epsilon]$ and using \eqref{phisol}, \eqref{fgensol} and \eqref{minusconst} we get 
\begin{equation}
    \label{gda2}
    4 \gamma_{+} \delta_{+} \alpha_{+} = (\gamma_{+} - \delta_{+})^2 Q -  |\gamma_{+} + \delta_{+}| (\gamma_{+} - \delta_{+})  |Q| \,.
\end{equation}
From the relations \eqref{gda1} and \eqref{gda2} it follows that 
\begin{equation}
    \label{gdQ}
    |\gamma_{+} + \delta_{+}| (\gamma_{+} - \delta_{+}) |Q| = (\gamma_{+}^2 + \delta_{+}^2) Q \, .
\end{equation}
Squaring the last equation we obtain 
\begin{equation}
\label{gd=0}
    \gamma_{+} \delta_{+} = 0 \, ,
\end{equation}
so exactly one of $\gamma_{+}$, $\delta_{+}$ is equal to zero, since $\gamma_{+}$ and $\delta_{+}$ cannot be both zero. Then, the relations \eqref{minusconst} imply that 
\begin{equation}
    \label{minus=plus}
    \beta_{-} = \beta_{+} \equiv \beta\,, \qquad \gamma_{-} = \gamma_{+} \equiv \gamma\,, \qquad \delta_{-} = \delta_{+} \equiv \delta  \,,
\end{equation}
while from \eqref{gdQ} and \eqref{gd=0} it follows that 
\begin{equation}
    \label{sgngdQ}
\sgn (\gamma - \delta) = \sgn Q\,.
\end{equation}
Moreover, we note that $\beta^2$ appears as a constant coefficient in $g_{tt} = - e^{2f}$ (see \eqref{fgensol}), so we can set $\beta^2 = 1$ without loss of generality. More precisely, we must take 
\begin{equation}
    \label{betavalue}
    \beta = \sgn (\gamma - \delta) = \sgn Q \,,
\end{equation}
so that the function $f$ is defined at $x=0$. 

If $Q \phi(0) >0$, then $\sgn (\gamma + \delta) = \sgn Q = \sgn(\gamma - \delta)$, which can be satisfied only if $\gamma \neq 0$ and $\delta =0$, given that exactly one of $\gamma$, $\delta$ is zero. In this case, $\gamma = \phi (0)$ and $\beta \gamma = (\sgn \gamma) \gamma = |\gamma| = |\phi (0)|$ and the ordinary differential equations \eqref{dilELans}-\eqref{Einsxxans} are solved by
\begin{align}
\label{phiQphi>0}
\phi(x) &= \phi (0) \exp \left[ \frac{1}{2} (\alpha + Q H(-x)) x \right]= \left\{ \begin{array}{ll}
     \phi(0) e^{\frac{1}{2} \alpha x} & x\geq 0  \\
     \phi(0) e^{\frac{1}{2} (\alpha +Q ) x} & x<0 
\end{array}   \right.  , \\
\label{fQphi>0}
f(x) &= \ln |\phi (x)| = \left\{ \begin{array}{ll}
   \ln |\phi(0)| + \frac{1}{2} \alpha x & x\geq 0 \\
   \ln |\phi(0)| + \frac{1}{2} (\alpha + Q) x & x < 0 
    \end{array}   \right. \,, \\
    \label{BQphi>0}
B (x) & = (\sgn Q )\phi^2 (0) \exp [(\alpha + Q H(-x))x]=    \left\{ \begin{array}{ll}   
    (\sgn Q )\phi^2 (0) e^{\alpha x} & x \geq 0 \\
    (\sgn Q )\phi^2 (0) e^{(\alpha + Q) x} & x<0
\end{array}   \right.  \,, 
    \end{align}
where we have set $\alpha_{+} = \alpha$ and $H$ is the Heaviside step function with $H(0)=1$. 

On the other hand, if $Q \phi(0) <0$, then $\sgn (\gamma + \delta) = -\sgn Q = -\sgn(\gamma - \delta)$, which can hold only if $\gamma = 0$ and $\delta  \neq 0$, given that exactly one of $\gamma$, $\delta$ is zero. In this case, $\delta = \phi (0)$ and $\beta \delta = - (\sgn \delta) \delta = - |\delta| = - |\phi(0)|$ and the solution to equations \eqref{dilELans}-\eqref{Einsxxans} is 
\begin{align}
\label{phiQphi<0}
\phi(x) &= \phi (0) \exp \left[- \frac{1}{2} (\alpha + Q H(-x)) x \right]= \left\{ \begin{array}{ll}
     \phi(0) e^{-\frac{1}{2} \alpha x} & x\geq 0  \\
     \phi(0) e^{-\frac{1}{2} (\alpha +Q ) x} & x<0 
\end{array}   \right.  , \\
\label{fQphi<0}
f(x) &= \ln |\phi (x)|= \left\{ \begin{array}{ll}
   \ln |\phi(0)| - \frac{1}{2} \alpha x & x\geq 0 \\
   \ln |\phi(0)| - \frac{1}{2} (\alpha + Q) x & x < 0 
    \end{array}   \right.  \,, \\
    \label{BQphi<0}
B (x) & = (\sgn Q )\phi^2 (0) \exp [-(\alpha + Q H(-x))x]  =    \left\{ \begin{array}{ll}   
    (\sgn Q )\phi^2 (0) e^{-\alpha x} & x \geq 0 \\
    (\sgn Q )\phi^2 (0) e^{-(\alpha + Q) x} & x<0
\end{array}   \right.\,.
    \end{align}
One can check that the solutions \eqref{phiQphi>0}-\eqref{BQphi>0} and \eqref{phiQphi<0}-\eqref{BQphi<0} to the differential equations \eqref{dilELans}-\eqref{Einsxxans} are valid even if $\alpha_{+} = \alpha =0$ or $\alpha_{-} = 0 \Leftrightarrow \alpha = - Q$.

\section{Dp-brane action and energy-momentum tensor }

In the Einstein frame, the action that describes a $Dp$-brane in 10D spacetime reads
\begin{equation}
    \label{SDp}
    S_{\rm{Dp}} = - T_p \int_{\mathcal{W}_{p+1}} d^{p+1} \xi \, e^{\frac{p-3}{4} \varphi} \sqrt{-g_{(\mathcal{W}_{p+1})}} + \mu_p \int_{\mathcal{W}_{p+1}} C_{p+1} \,,
\end{equation}
where $\mathcal{W}_{p+1}$ denotes the brane worldvolume, which is parametrized by the coordinates $\xi^{\alpha}$, $\alpha = 0,1,\dots\,,p$. Its embedding into 10D spacetime is described by parametric equations of the form $x^M = X^M(\xi^{\alpha})$. Also, $g_{(\mathcal{W}_{p+1})}$ is the determinant of the induced metric on $\mathcal{W}_{p+1}$, namely 
\begin{equation}
    \label{gindDp}
    g_{\alpha \beta}^{(\mathcal{W}_{p+1})} = g_{MN} (X(\xi)) \frac{\partial X^M}{\partial \xi^{\alpha}} \frac{\partial X^N}{ \partial \xi^{\beta}
    }.
\end{equation}
Furthermore, $T_p = \frac{1}{(2\pi)^p l_s^{p+1}}$ and $\mu_{p}$ is the charge of the $Dp$-brane under the RR (p+1)-form $C_{p+1}$. The energy momentum tensor associated with the $Dp$-brane described by \eqref{SDp} reads 
\begin{equation}
\begin{aligned}
    \label{TDpMN}
    T^{\rm{Dp}}_{MN} \equiv & - \frac{2}{\sqrt{-g_{(10)}}} \frac{\delta S_{ \rm{Dp}}}{\delta g^{MN}}  \\
    =&  -\frac{1}{\sqrt{-g_{(10)}}} T_p e^{\frac{p-3}{4} \varphi} g_{MP} g_{NQ} \int_{\mathcal{W}_{p+1}} d^{p+1} \xi \sqrt{-g_{(\mathcal{W}_{p+1})}} \, g^{\alpha \beta}_{(\mathcal{W}_{p+1})} \frac{\partial X^P}{\partial \xi^{
    \alpha
    }} \frac{\partial X^Q}{\partial \xi^{
    \beta
    }} \delta^{(10)} (x-X(\xi)) \,,
 \end{aligned}
 \end{equation}
where $g_{(10)} = \det g_{MN}$ and $g^{\alpha \beta}_{(\mathcal{W}_{p+1})}$ is the inverse of the induced metric on $\mathcal{W}_{p+1}$.

\bibliographystyle{utphys}
\bibliography{references}

\end{document}